\documentclass[useAMS,usenatbib]{mn2e}

\pdfoutput=1 
\usepackage{psfig} 
\usepackage{amssymb} 
\usepackage[pdftex]{graphicx}

\def\btg{G\"{a}nsicke}
\def\galex{{\em GALEX}}
\def\halpha{H$\alpha$}
\def\hbeta{H$\beta$}
\def\hgamma{H$\gamma$}
\def\hei{He{\sc i}}
\def\heil{He{\sc i}$\;\lambda$}
\def\heii{He{\sc ii}}
\def\heiil{He{\sc ii}$\;\lambda$}
\def\caii{Ca{\sc ii}}
\def\Porb{P_{\mbox{\small orb}}}

\title[1000 CVs from CRTS]{One thousand cataclysmic variables from the Catalina Real-time Transient Survey}

\author[E. Breedt et al.]{E.~Breedt$^1$\thanks{E-mail: e.breedt@warwick.ac.uk}, 
  B.\,T.~\btg$^1$, A.\,J.~Drake$^2$, P.~Rodr\'{i}guez-Gil$^{3,4}$, S.\,G.~Parsons$^5$, \and T.\,R.~Marsh$^1$, P.~Szkody$^6$, M.\,R.~Schreiber$^5$, S.\,G.~Djorgovski$^2$\\
$^1$ Department of Physics, University of Warwick, Coventry, UK\\
$^2$ California Institute of Technology, 1200 E. California Blvd, CA 91225, USA\\
$^3$ Instituto de Astrof\'\i sica de Canarias, V\'\i a L\'actea s/n, La Laguna, E-38205, Santa Cruz de Tenerife, Spain\\
$^4$ Departamento de Astrof\'\i sica, Universidad de La Laguna, La Laguna, E-38206, Santa Cruz de Tenerife, Spain\\
$^5$ Instituto de F\'isica y Astronom\'ia, Universidad de Valpara\'iso, Avenida Gran Bretana 1111, Valpara\'iso, Chile \\
$^6$ Department of Astronomy, University of Washington, Box 351580, Seattle, WA 98195-1580, USA\\ 
}

\date{Accepted/Received}

\pagerange{\pageref{firstpage}--\pageref{lastpage}} \pubyear{2014}

\begin{document}

\label{firstpage}

\maketitle

\begin{abstract}
Over six years of operation, the Catalina Real-time Transient Survey (CRTS) has identified 1043 cataclysmic variable (CV) candidates --- the largest sample of CVs from a single survey to date. Here we provide spectroscopic identification of 85 systems fainter than $g\geq19$, including three AM\,CVn binaries, one helium-enriched CV, one polar and one new eclipsing CV. 
We analyse the outburst properties of the full sample and show that it contains a large fraction of low accretion rate CVs with long outburst recurrence times. We argue that most of the high accretion rate dwarf novae in the survey footprint have already been found and that future CRTS discoveries will be mostly low accretion rate systems. We find that CVs with white dwarf dominated spectra have significantly fewer outbursts in their CRTS light curves compared to disc-dominated CVs, reflecting the difference in their accretion rates. 
Comparing the CRTS sample to other samples of CVs, we estimate the overall external completeness to be 23.6 per cent, but show that as much as 56 per cent of CVs have variability amplitudes that are too small to be selected using the transient selection criteria employed by current ground-based surveys. 
The full table of CRTS CVs, including their outburst and spectroscopic properties examined in this paper, is provided in the online materials.

\end{abstract}



%
\section{Introduction} \label{sec:intro}

Cataclysmic Variable stars (CVs) are compact binary stars consisting of white dwarfs accreting from Roche lobe filling, low mass, main sequence stars. (See \citealt{warnerbook} for a comprehensive review.)
The accretion is driven by angular momentum loss, and the process is equally applicable to the long term evolution of a variety of other compact binaries such as millisecond pulsars \citep[e.g.][]{taam00} and X-ray binaries \citep[e.g.][]{vanhaaften13}. As accreting white dwarfs, CVs also remain possible progenitors of Type Ia supernovae \citep{zorotovic11,kafka12,toonen14}, which play a central role in calibrating the extragalactic distance scale. Compared to most other compact binary populations, CVs are plentiful, bright and nearby, making them the ideal population for the development and testing of binary evolution models \citep[e.g][]{kolb93,howell97,howellnr01}.

In the majority of CVs, accretion occurs via an accretion disc around the white dwarf. The disc often dominates the luminosity of the system, so the component stars are not visible in the optical part of CV spectra unless the accretion rate is sufficiently low and the disc correspondingly faint. 
The accretion discs in CVs are subject to a thermal instability, which causes quasi-periodic, large-amplitude outbursts, known as ``dwarf nova outbursts'' \citep{mmh81,osaki89}. The onset of the outburst is fast: the system brightens by 2 -- 6 magnitudes over a period as short as a day. The disc typically stays in a hot, viscous, bright state for a few days, then fades back to its quiescent brightness.
Outburst recurrence times vary from system to system and depend primarily on the mass transfer rate from the donor star and the size of the accretion disc. At the two extremes, the ER\,UMa-type stars are seen in outburst every few days while the WZ\,Sge-type stars can spend several decades in quiescence before another outburst occurs \citep{robertson95,patterson02}. 
As a result of these bright outbursts, dwarf novae are commonly observed as optical transients, and variability remains one of the most efficient ways of discovering new CVs. 
Currently active large area transient surveys include the Catalina Real-time Transient Survey \citep[CRTS,][]{drake09crts}, the Palomar Transient
Factory \citep[PTF,][]{ptf}, the Mobile Astronomical System of the Telescope Robots \citep[MASTER,][]{masterrobotic} and the All Sky Automated Survey \citep[ASAS,][]{asas}. CVs are also expected to be a significant fraction of the transients detected by future surveys such as  the Gaia Science Alerts \citep{gaia} and the Large Synoptic Survey Telescope survey \citep[LSST,][]{lsst}.

Most of the dwarf novae with orbital periods shorter than 2~hours also show superoutbursts, which last longer and are brighter than normal outbursts. During superoutbursts, the light curves of these CVs display photometric modulations on a period slightly longer than the orbital period. These are known as `superhumps' and result from the dynamical interaction between the disc and the donor star.  The superhump period is closely related to the mass ratio of the two component stars and is a good proxy for the orbital period \citep[e.g.][]{patterson05,katoI09,gaensicke09}.

The orbital period is the most readily observable property of a CV, and provides important clues to the evolutionary state of the system. CVs evolve from long (few hours) to short periods, down to a minimum of $\sim80$~min. At this point, the donor has lost so much mass that it cannot sustain hydrogen burning in its core any longer and it becomes degenerate. The thermal timescale of the donor star (i.e. the timescale on which it responds to the mass loss) becomes longer than the mass loss timescale, so it cannot contract rapidly enough to establish thermal equilibrium. In order to  accommodate the star, the binary orbit has to expand and the orbital period increases again \citep[e.g.][]{rjw82,kolbbaraffe99}. As a system evolves towards the period minimum, the accretion rate drops and the evolution slows down, so a key prediction of these evolutionary models is an accumulation of CVs at the minimum period. Despite it being a long-standing prediction, the first observational confirmation of such a spike in the orbital period histogram was only found recently, in the Sloan Digital Sky Survey (SDSS) sample of CVs \citep{gaensicke09}. There are two main reasons for the success of the SDSS-based work in uncovering the previously ``missing'' CVs. Firstly, by selecting targets spectroscopically, a well-known observational bias could be eliminated. CVs near the period minimum have very low accretion rates, so undergo less frequent outbursts. A survey which selects CVs based on large amplitude variability, such as dwarf nova outbursts, is therefore biased against these WZ\,Sge type systems, which have long outburst recurrence times, but which make up a considerable, if not dominant, fraction of the entire CV population \citep[e.g.][]{pretorius07a,howell01}. The low accretion rate CVs are also intrinsically faint, so the unprecedented depth of the SDSS spectroscopic survey ($g\lesssim19.0$) was the second key ingredient needed to find the period minimum spike. 
Despite the bias towards CVs with infrequent outbursts, transient surveys still play an important role in CV discovery, as they can probe much deeper over a large sky area than is possible with spectroscopic surveys. The bias can also be reduced by using a low threshold for transient selection, so that outbursts are detected even if the target is not at maximum brightness when it is observed, or that small amplitude variations are detected as well. The majority of CVs are photometrically variable, even if they do not display dwarf nova outbursts.

Our analysis in this paper is based on the Catalina Real-time Transient Survey (CRTS), which utilises the images from the Catalina Sky Survey to identify supernovae, CVs, flaring AGN and other optical transient phenomena \citep{drake09crts}. More than 9000 transient sources have been detected since the start of CRTS operations in 2007 November.
The Catalina Sky Survey itself started in 2004, with the main aim of finding near-earth objects\footnote{http://www.lpl.arizona.edu/css/}. It consists of three separate telescopes which repeatedly scan a combined $\sim30\,000$~deg$^2$ of sky between $-75\degr<\delta<65\degr$, except for a few degrees around the Galactic plane. The three telescopes run separate sub-surveys, known as the Catalina Schmidt Survey (CSS), the Mount Lemmon Survey (MLS) and Siding Spring Survey (SSS). Details of the telescopes and the observing program are given by \citet{drake09crts} and \citet{drake14}.
In terms of finding CVs, the Catalina Real-time Transient Survey benefits not only from a large sky area and an increased depth compared to other transient surveys, but also from the long baseline of observations. With a typical observing cadence of $\sim2$ weeks over the 6-year baseline, it has a higher probability of observing the infrequent outbursts of low accretion rate CVs. The limiting magnitude of the survey is 19 -- 21 (depending on the telescope), which is up to 2~mag deeper than the spectroscopic selection limit of SDSS.  Considering also the large amplitudes of dwarf nova outbursts, CRTS is, in principle, sensitive to dwarf novae with quiescent magnitudes as faint as $\sim$26. 

The survey has already discovered many new CVs, and is a rich source for follow-up studies \citep[e.g.][]{woudt12,thorstensenskinner12,coppejans13}.
The transient selection criteria, CV identification process, spatial distribution and photometric properties of the overall sample were presented by \citet{drake14}. In this paper, we focus on the spectroscopic and outburst properties of the sample. We first present identification spectroscopy of 85 systems --- 72 from our own follow-up observations, seven spectra from the latest data release of the Sloan Digital Sky Survey and a further six spectra from the Public ESO Spectroscopic Survey for Transient Objects. In the second part of this paper, we investigate the outburst properties of the full sample of CRTS CVs candidates (from CSS, MLS and SSS) up to and including 2013 November 11, a total of 1043 unique targets. We finally compare the sample to other known CV samples to make an estimate of the CRTS completeness and its CV detection efficiency. 

CRTS transients are assigned a detection ID in the form \{CSS/MLS/SSS\}\,{\em\{date\}}:HHMMSS$\pm$DDMMSS, to indicate the survey in which the target was first found, the discovery date (usually the first bright detection compared to previous observations and/or catalogues), and the target coordinates. Once the transient has been classified \citep[e.g.][]{drake09crts,drake14} it is assigned a CRTS identifier of the form CRTS\,JHHMMSS.S+DDMMSS. We will use the latter form to refer to the CVs in this paper, but include the survey telescope and discovery date information in the online table, as it is used in parts of our analysis. For simplicity, we will use an abbreviated form of the CRTS ID in figure labels, i.e. CRTS\,HHMM$\pm$DDMM.


%
\section{Identification spectroscopy} \label{sec:data}

We first present the spectra from our own observations in this section and the next. A further 13 spectra, identified in public spectroscopic surveys, will be presented in Section~\ref{sec:otherspec}.   

We obtained identification spectra of a total of 72 faint CV candidates identified by the CRTS, using the Gran Telescopio Canarias (GTC; La Palma, Spain) and the Gemini telescopes (North: Mauna Kea, Hawaii and South: Cerro Pachon, Chile). The observations were carried out in service mode during 2010, 2011 and 2013. A log of the observations is shown in Table~\ref{tab:obslog} and the spectra are shown in Figure~\ref{fig:spectra} in the appendix. We selected our targets to be fainter than $g=19$ and to fall within the SDSS photometric survey area, but made no selection based on the colour or the observed outburst activity.  The SDSS photometry provides a useful additional calibration of the Gemini spectra, as described below.

\begin{table*}
\caption{\label{tab:obslog} 
Observation log of our spectroscopic identification program, ordered by target right ascension. The Gemini spectra consist of three individual exposures. The observation date refers only to the first spectrum of that target, but the exposure time is the total exposure time in minutes. (This is an extract only. The complete table listing all 72~targets is available in the supplementary materials online.) }
\begin{tabular}{crrclcl}
\hline
  \multicolumn{1}{c}{} &
  \multicolumn{1}{c}{RA} &
  \multicolumn{1}{c}{DEC} &
  \multicolumn{1}{c}{} &
  \multicolumn{1}{c}{OBSERVATION} &
  \multicolumn{1}{c}{EXP.} &
  \multicolumn{1}{c}{} \\
  \multicolumn{1}{c}{CRTS ID} &
  \multicolumn{1}{c}{(deg)} &
  \multicolumn{1}{c}{(deg)} &
  \multicolumn{1}{c}{INSTRUMENT} &
  \multicolumn{1}{c}{DATE (UT)} &
  \multicolumn{1}{c}{(min)} &
  \multicolumn{1}{c}{COMMENT} \\
\hline\hline
  CRTS$\:$J004902.0$+$074726  &  $12.259$ & $  7.791$ & GMOS-South & 2010-09-14\hspace{2mm} 06:41:06.9 & 46.0 & PA $>70$\degr\, off parallactic\\
  CRTS$\:$J010522.2$+$110253  &  $16.342$ & $ 11.048$ & GMOS-South & 2010-12-04\hspace{2mm} 00:53:20.8 & 42.0 &  \\
  CRTS$\:$J011306.7$+$215250  &  $18.278$ & $ 21.881$ & GMOS-North & 2010-09-18\hspace{2mm} 13:00:27.8 & 45.0 &  \\
  CRTS$\:$J035003.4$+$370052  &  $57.514$ & $ 37.014$ & GMOS-North & 2010-09-19\hspace{2mm} 13:50:37.8 & 27.0 &  \\
  CRTS$\:$J043517.8$+$002941  &  $68.824$ & $  0.495$ & GMOS-South & 2010-12-15\hspace{2mm} 05:27:56.7 & 46.0 &  \\
  CRTS$\:$J043546.9$+$090837  &  $68.945$ & $  9.144$ & GMOS-South & 2010-11-05\hspace{2mm} 03:23:17.0 & 47.0 & Poor seeing, high humidity \\
  CRTS$\:$J043829.1$+$004016  &  $69.621$ & $  0.671$ & GMOS-South & 2010-12-04\hspace{2mm} 07:23:01.7 & 28.0 &  \\
  CRTS$\:$J073339.3$+$212201  & $113.414$ & $ 21.367$ & GMOS-North & 2010-11-12\hspace{2mm} 12:19:00.8 & 45.0 & Cloudy\\
  \dots & \dots & \dots & \dots & \dots & \dots &  \\                                                                                             
\hline
\end{tabular}                                                                                       
\end{table*}

\subsection{Gemini GMOS spectroscopy} \label{sec:gemini}

All the Gemini observations were made in Service Mode, using the Gemini Multi-Object Spectrographs (GMOS) fitted with a low resolution grating (R150) and a 1\arcsec\, slit. Each observation was executed as three individual exposures: one at a central wavelength setting of 7400\AA\, and the other two with a central wavelength shift of 50\AA\, up and down. This was done to (partially) fill the gaps in the spectrum caused by the physical gaps between the three CCDs of the detector array, and also to improve the background subtraction. 

The observations were carried out between 2010 August and 2011 May, and 2013 February to June. The 2010 and 2013 observations were carried out as Band-3 programmes, meaning that weather conditions were often cloudy and/or the seeing poor. We compensated for this by using longer exposure times. Occasionally, if the conditions deteriorated too much, the three exposures could not be carried out consecutively. The balance was then observed later in the night when the weather had sufficiently improved, or on a following night. The date listed in Table~\ref{tab:obslog} corresponds to the first observation for each target. 
The 2013 observations benefited from an upgrade to the GMOS-North detector array to a set of deep depletion CCDs with greater sensitivity and less fringing. This led to much improved spectra, despite the Band-3 weather conditions. For these observations we used two central wavelength settings, 7400\AA\, and 7350\AA, as well as nodding along the slit to improve the sky subtraction. 

Neither GMOS is equipped with an Atmospheric Dispersion Corrector (ADC), so a few of our observations were affected by flux loss in the blue part of the spectrum. The effect is usually minimised by observing at the parallactic angle, i.e. aligning the slit with the direction of the dispersion, so that most of the flux enters the slit. The parallactic angle changes with the target's position on the sky, so it cannot be predetermined for queue-scheduled observations like these. Unfortunately, on GMOS the slit angle is set by the choice of guide star for the on-instrument wavefront sensor. In Band-3 conditions, this guide star needs to be bright enough to enable guiding through clouds, limiting the number of suitable guide stars, and hence the number of slit position angles available for the observation.  
In order to minimise slit losses, we identified multiple guide stars where possible and used a wide slit. We also selected the $g$ band filter for target acquisition, which means that the target was centred on the slit near the middle of the spectral range. 
We chose not to impose airmass restrictions on our observations, in order to increase the scheduling flexibility of the programme. Occasionally this meant that observations had to be carried out using a slit angle far off the parallactic angle, with a detrimental effect on the blue flux. The targets affected by this problem are marked with asterisks in Figure~\ref{fig:spectra}.

The acquisition images from GMOS are written to disc and are available as part of the data package. We carried out photometry on these images, but because of the Band-3 weather conditions, the measurements are not very accurate. However, since a dwarf nova outburst typically causes a brightening of several magnitudes, we could reliably use the photometry to determine whether an object was in outburst at the time of our observations.

We reduced and optimally extracted the spectra from CCD2 and CCD3, using {\sc starlink} packages {\sc figaro}, {\sc kappa} and {\sc pamela}. CCD1 covers the spectral region redwards of 9300\AA, where almost all of our targets were undetected for the exposure times used, so it was discarded. Significant fringing is apparent in the extracted spectra redwards of about 7000\AA, especially for the faintest targets. The wavelength calibration is based on copper-argon arc lamp exposures taken during the day as part of the standard calibration program. The wavelength calibration for CCD3 (the bluest part of the spectrum) is rather poor, because of the small number of arc- and sky lines are found in this spectral range (rms of the residuals for the blue wavelength fit is $0.33$\AA, compared to $0.08$\AA\, for the red part of the spectrum).  We therefore adjusted the wavelength scale to shift the emission lines to their zero velocity position before co-adding the individual exposures. This has the added advantage of reducing the amount of orbital smearing of the emission lines. The extracted spectra have a spectral range of $\sim3800-9200$\AA, with a resolution of 20.7\AA, as measured from width of night sky emission lines. 

The spectroscopic flux calibration was applied in two steps. We first corrected for the instrument response using a standard star observation taken during each observing semester, then we scaled each spectrum by a constant factor to match its quiescent SDSS photometry\footnote{CVs are inherently variable objects, so multiple photometric measurements often differ. We considered all available photometric measurements from SDSS DR7 and DR9, and took the faintest set of $ugriz$ photometry as a best estimate of the quiescent magnitude of each target. The $g$ band magnitudes determined this way are shown in Table~\ref{tab:obslog}.}. If the target was in outburst during our observations, or considerably fainter than the assumed quiescent brightness from SDSS, we applied no scaling. Flux and colour differences between our spectra and the photometry are obvious in several targets (Figure~\ref{fig:spectra}), but that is expected given how variable CVs are. The wavelength and flux calibration, and averaging of the individual exposures were done using {\sc molly}\footnote{{\sc molly} was written by TRM and is available from http://www.warwick.ac.uk/go/trmarsh/software/.}.

\subsection{GTC OSIRIS spectroscopy}
Seven targets were observed with the OSIRIS spectrograph on the Gran Telescopio Canarias (also known as GTC or GranTeCan) in 2011 March and May (see Table~\ref{tab:obslog} online and Figure~\ref{fig:spectra}). 
The observations were carried out in Service Mode, using the R300B grism, a 1\arcsec\, wide slit and $2\times2$ binning of the CCD. This setup delivers a spectral range of approximately 3900--9970\AA\, at a resolution of 11.6\AA. Acquisition was done in the $r$ band using 15 second exposures and all spectra were taken at the parallactic angle, under spectroscopic sky conditions. Exposure times for the individual targets varied between 2 -- 30 minutes. 

We debiased and flat fielded the spectra using standard procedures in IRAF\footnote{IRAF is distributed by the National Optical Astronomy Observatory, which is operated by the Association of Universities for Research in Astronomy (AURA) under cooperative agreement with the National Science Foundation.}. As for the Gemini spectra above, we used {\sc pamela} to optimally extract the spectra and {\sc molly} to carry out the wavelength calibration. Finally we flux calibrated the spectra using observations of a standard star, GD140, taken on the same night as the target observations.


%
\section{Notes on individual objects} \label{sec:targetnotes}

All 72 of the targets we observed have spectra that are typical of accreting white dwarf binary systems. The spectra are shown in Figure~\ref{fig:spectra}, arranged by increasing right ascension. In this section, we use the spectra and the corresponding CRTS light curves to classify the targets, and highlight a few systems worthy of further observations. The main spectral properties are summarised in Table~\ref{tab:targetinfo}.

\begin{table*}
\caption{\label{tab:targetinfo} Spectral properties and classification of our spectroscopic targets. The $g$ band magnitudes listed indicate our best estimates of the quiescent brightness of each target and were taken from SDSS DR7 and DR9. The velocity width refers to the H$\alpha$ emission line, or in the case of AM\,CVn systems, the width of the \heil5876 line. No value is shown for outburst spectra. All periods were taken from literature, as indicated in the References column. A colon following the period value indicates that the period is uncertain. The spectral classification symbols are as follows: DN = dwarf nova, SU = SU\,UMa star, W = white dwarf visible in the spectrum, S = Secondary star visible in the spectrum, He = AM\,CVn star, P = polar, uHe = unusual spectrum, Hi = high inclination, inferred from double-peaked emission lines and/or linewidth, Ca = \caii\, emission, O = outburst spectrum, E = confirmed eclipser. }

\begin{tabular}{cccccccc}
\hline
  \multicolumn{1}{c}{} &
  \multicolumn{1}{c}{} &
  \multicolumn{1}{c}{SDSS} &
  \multicolumn{1}{c}{FWHM$^\dagger$} &
  \multicolumn{1}{c}{$\Porb$} &
  \multicolumn{1}{c}{} &
  \multicolumn{1}{c}{CLASSIFICATION/} \\
  \multicolumn{1}{c}{CRTS ID} &
  \multicolumn{1}{c}{ALT. NAME} &
  \multicolumn{1}{c}{$g$ mag} &
  \multicolumn{1}{c}{(km/s)} &
  \multicolumn{1}{c}{(d)} &
  \multicolumn{1}{c}{REF.$^\P$} &
  \multicolumn{1}{c}{PROPERTIES} \\
\hline\hline
  CRTS$\:$J004902.0$+$074726 &             &  21.52 & 2038    & $-$                  & $-$ & Hi   & \\
  CRTS$\:$J010522.2$+$110253 &             &  20.95 & 1427    & $-$                  & $a$ & W, S & \\
  CRTS$\:$J011306.7$+$215250 &  GV\,Psc    &  20.43 &  799    & 0.0902(6)$^\ddagger$ &$a,b$& SU, Ca  & \\
  CRTS$\:$J035003.4$+$370052 &             &  19.31 & 1688    & $-$                  & $-$ & DN, Hi  & \\
  CRTS$\:$J043517.8$+$002941 & PTF1\,J043517.73$+$002940.7 &  21.97 & 2198$^*$& 0.0238(13)  & $c$ & He  & \\
  CRTS$\:$J043546.9$+$090837 &             &  21.84 & 1039    & $-$                  & $-$ & W   & \\
  CRTS$\:$J043829.1$+$004016 &             &  19.72 & 2078    & 0.06546(9)           & $d$ & Hi, E & \\
  CRTS$\:$J073339.3$+$212201 &             &  20.45 & 1129    & $-$                  & $-$ & S   & \\
  CRTS$\:$J074419.7$+$325448 &             &  21.31 & 3324$^*$& $-$                  & $-$ & He  &\\
  CRTS$\:$J074928.0$+$190452 &             &  21.14 & 1222    & $-$                  & $-$ & DN  & \\
  CRTS$\:$J075332.0$+$375801 &             &  20.16 & 1374    & $-$                  & $-$ & DN, Ca  & \\
  CRTS$\:$J075414.5$+$313216 &             &  20.01 & 1683    & 0.0615(5)$^\ddagger$ & $e$ & SU, W, Ca & \\
  CRTS$\:$J080853.7$+$355053 &             &  19.61 & 1911    & $-$                  & $-$ & uHe & \\
  CRTS$\:$J081414.9$+$080450 &             &  24.27 &  $-$    & $-$                  & $-$ & DN, O & \\ 
  CRTS$\:$J081936.1$+$191540 &             &  20.38 & 1158    & $-$                  & $-$ & S   & \\
  CRTS$\:$J082019.4$+$474732 &             &  21.37 & 1777    & $-$                  & $-$ & DN, Ca  & \\
  CRTS$\:$J082123.7$+$454135 &             &  20.60 & 1367    & $-$                  & $a$ & S   & \\
  CRTS$\:$J082654.7$-$000733 &             &  19.54 & 2203    & 0.05976(8)           & $f$ & Hi, E, W & \\
  CRTS$\:$J082821.8$+$105344 &             &  22.34 & 1662    & $-$                  & $-$ & W   & \\
  CRTS$\:$J084041.5$+$000520 &             &  20.81 & 1060    & $-$                  & $-$ & DN  & \\
  CRTS$\:$J084108.1$+$102537 &             &  21.13 &  984    & $-$                  & $-$ & DN, Ca  & \\
  CRTS$\:$J084127.4$+$210054 &             &  20.61 & 2280    & 0.0841(6)$^\ddagger$ & $g$ & SU, Hi, W, S & \\
  CRTS$\:$J084358.1$+$425037 &             &  19.91 & 1031    & $-$                  & $-$ & DN, Ca  & \\
  CRTS$\:$J084413.6$-$012807 &             &  20.54 & 2205$^*$& $-$                  & $-$ & He  &\\
  CRTS$\:$J084555.0$+$033929 &  V498\,Hya  &  20.87 & 1518    & 0.05915(48)$^\ddagger$&$b$ & SU, W   & \\
  CRTS$\:$J085113.4$+$344449 &             &  20.44 &  944    & 0.077:$^\ddagger$    & $i$ & SU, W, S, Ca & \\
  CRTS$\:$J085409.4$+$201339 &             &  20.90 & 1828    & $-$                  & $-$ & DN  & \\
  CRTS$\:$J085603.8$+$322109 &             &  19.64 & 2183    & $-$                  & $-$ & Hi, W?   & \\
  CRTS$\:$J090016.7$+$343928 &             &  20.46 &  912    & $-$                  & $-$ & DN, uHe & \\
  CRTS$\:$J090239.7$+$052501 &             &  23.15 & 2007    & 0.0555(5)$^\ddagger$ &$b,h$& SU, Hi, E? & \\
  CRTS$\:$J090516.0$+$120451 &             &  19.80 & 1221    & $-$                  & $a$ & DN  & \\
  CRTS$\:$J091634.6$+$130358 &             &  22.03 & 2015    & $-$                  & $-$ & DN, Hi  & \\
  CRTS$\:$J101545.9$+$033312 &             &  20.18 &  820    & $-$                  & $-$ & DN, Ca  & \\
  CRTS$\:$J102616.0$+$192045 &             &  20.12 & 1658    & 0.0796(6)$^\ddagger$ & $e$ & SU, W, S, Ca & \\
  CRTS$\:$J102637.0$+$475426 &             &  20.16 & 1943    & 0.0667(5)$^\ddagger$ & $j$ & SU, Hi, W, Ca   & \\
  CRTS$\:$J102937.7$+$414046 &             &  22.39 & 1705    & $-$                  & $-$ & DN  & \\
  CRTS$\:$J103317.3$+$072119 &             &  19.89 & 1521    & $-$                  & $-$ & W, S, Ca & \\
  CRTS$\:$J104411.4$+$211307 &             &  19.36 & 1821    & 0.0589(5)$^\ddagger$ & $j$ & SU, W, Ca   & \\
  CRTS$\:$J112332.0$+$431717 &             &  19.92 & 1800    & $-$                  & $-$ & DN  & \\
  CRTS$\:$J112509.7$+$231036 &             &  20.99 & 2099    & $-$                  & $-$ & DN, Hi, Ca  & \\
  CRTS$\:$J113708.6$+$513451 &             &  20.67 &  898    & $-$                  & $-$ & DN  & \\
  CRTS$\:$J113950.6$+$455818 &  NSV\,5285  &  19.63 & 1997    & 0.0843(6)$^\ddagger$ &$a,b$& SU, W, S  & \\
  CRTS$\:$J115330.2$+$315836 & SDSS\,J115330.25$+$315835.9 &  20.12 & 1957    & $-$  & $-$ & DN  & \\
  CRTS$\:$J124819.4$+$072050 &             &  21.46 & 2318    & $-$                  & $-$ & DN, Hi  & \\
  CRTS$\:$J130030.3$+$115101 &             &  19.83 & 1965    & 0.06267(8)           &$b,f$& Hi, W, S & \\
  CRTS$\:$J132103.2$+$015329 &  HV\,Vir    &  19.23 & 1748    & 0.057069(6)          &$f,k$& W & \\
  CRTS$\:$J132536.0$+$210037 &             &  23.03 & 1729    & $-$                  & $r$ & Hi, E, W, Ca & \\
  CRTS$\:$J135219.0$+$280917 &             &  20.69 & 1748    & $-$                  & $-$ & S & \\
  CRTS$\:$J140454.0$-$102702 &             &  19.76 & 2157    & 0.059579(7)          & $f$ & Hi, E, W, S? & \\
  CRTS$\:$J141712.0$-$180328 &             &  20.61 & 1838    & 0.0845(1)            & $d$ & DN, W, S, Ca  & \\
  CRTS$\:$J144011.0$+$494734 &             &  21.45 & 1954    & 0.0629(5)$^\ddagger$ & $j$ & SU, W & \\
  CRTS$\:$J145502.2$+$143815 &             &  20.13 & 1256    & $-$                  & $-$ & S & \\
  CRTS$\:$J145921.8$+$354806 &             &  21.55 & 1073    & 0.0817(6)$^\ddagger$ & $l$ & SU, W, S & \\
  \dots                     & \dots                       & \dots  & \dots   & \dots                &\dots& \dots & \\   
\end{tabular}
\end{table*}

\addtocounter{table}{-1}
\begin{table*}
\caption{
{\em continued.} Spectral properties and classification of our spectroscopic targets.}
\begin{tabular}{cccccccc}
\hline
  \multicolumn{1}{c}{} &
  \multicolumn{1}{c}{} &
  \multicolumn{1}{c}{SDSS} &
  \multicolumn{1}{c}{FWHM$^\dagger$} &
  \multicolumn{1}{c}{$\Porb$} &
  \multicolumn{1}{c}{} &
  \multicolumn{1}{c}{CLASSIFICATION/} \\
  \multicolumn{1}{c}{CRTS ID} &
  \multicolumn{1}{c}{ALT. NAME} &
  \multicolumn{1}{c}{$g$ mag} &
  \multicolumn{1}{c}{(km/s)} &
  \multicolumn{1}{c}{(d)} &
  \multicolumn{1}{c}{REF.$^\P$} &
  \multicolumn{1}{c}{PROPERTIES} \\
\hline\hline
  \dots                     & \dots                       & \dots  & \dots   & \dots                 &\dots&\dots& \\   
  CRTS$\:$J151020.7$+$182303 & \phantom{PTF1\,J043517.73$+$002940.7}           &  21.38 & 1877    & $-$                   & $-$ & S   & \\
  CRTS$\:$J152158.8$+$261223 &            &  21.09 & 1993    & $-$                   & $-$ & Hi, W & \\
  CRTS$\:$J154544.9$+$442830 &            &  20.95 & 1171    & 0.0742(5)$^\ddagger$  & $e$ & DN, Ca  & \\
  CRTS$\:$J155430.6$+$365043 &            &  21.74 & 1791    & 0.069:                & $m$ & DN  & \\
  CRTS$\:$J160003.7$+$331114 &  VW\,CrB   &  19.89 & 2106    & 0.0705(5)$^\ddagger$  & $n$ & Hi, W, S   & \\
  CRTS$\:$J160844.8$+$220610 &            &  20.94 & 1845    & $-$                   & $-$ & DN, Hi  & \\
  CRTS$\:$J163805.4$+$083758 &  V544\,Her &  19.74 & 1258    & 0.069(3):             & $p$ & S   & \\
  CRTS$\:$J164748.0$+$433845 &            &  21.59 & 3651    & $-$                   & $-$ & uHe, Hi & \\
  CRTS$\:$J165002.8$+$435616 &            &  22.91 & 1751    & $-$                   & $-$ & DN  & \\
  CRTS$\:$J171202.9$+$275411 &            &  21.42 &  958    & $-$                   & $-$ & DN  & \\
  CRTS$\:$J171223.1$+$362516 &            &  20.87 &  $-$    & $-$                   & $-$ & DN, O & \\ 
  CRTS$\:$J172515.5$+$073249 &            &  20.66 & 1736    & $-$                   & $-$ & W   & \\
  CRTS$\:$J173307.9$+$300635 &            &  22.53 & 1378    & $-$                   & $-$ & DN, S?  & \\
  CRTS$\:$J210043.9$-$005212 &            &  22.27 &  $-$    & $-$                   & $-$ & DN, O & \\ 
  CRTS$\:$J210650.6$+$110250 &            &  20.51 & 1556    & $-$                   & $-$ & DN, Ca  & \\
  CRTS$\:$J210954.1$+$163052 &            &  23.48 &  $-$    & $-$                   & $a$ & DN, O & \\ 
  CRTS$\:$J214842.5$-$000723 &  V340\,Aqr &  23.05 & 1607    & $-$                   & $-$ & DN  & \\
  CRTS$\:$J224823.7$-$092059 &            &  21.15 & 1628    & $-$                   & $-$ & W   & \\
  CRTS$\:$J231142.8$+$204036 &            &  21.24 &  996    & 0.08:                 & $q$ & DN, Ca  & \\
\hline                                                                                                 
\end{tabular}                                                                                          
\begin{footnotesize}                                                                                                      
\begin{flushleft}
$\dagger$ Refers to \halpha, except for the starred values, which refer to \heil5876.\\
$*$No hydrogen in the spectrum. Linewidth refers to the \heil5876 line.\\
$\ddagger$ Derived from the superhump period, using equation~1 of \citet{gaensicke09} \\
$\P$ References: 
  $a$ -- \citet{thorstensenskinner12}, 
  $b$ -- \citet{katoI09}, 
  $c$ -- \citet{levitan13}, 
  $d$ -- \citet{coppejans13}, 
  $e$ -- \citet{katoIII12}, 
  $f$ -- \citet{woudt12},
  $g$ -- \citet{katoV13},
  $h$ -- Uemura \& Arai (2008), vsnet-alert 9955,
  $i$ -- Maehara (2008), vsnet-alert 10723,
  $j$ -- \citet{katoII10},
  $k$ -- \citet{patterson03},
  $l$ -- \citet{katoIV13},
  $m$ -- Boyd (2008), vsnet-outburst 9500,
  $n$ -- \citet{nogami04},
  $p$ -- \citet{howell90},
  $q$ -- Masi (2012), vsnet-alert 14696,
  $r$ -- This paper.
\end{flushleft}
\end{footnotesize}
\end{table*}

\subsection{AM\,CVn systems and helium enriched CVs} \label{sec:amcvn}
AM Canum Venaticorum (AM\,CVn) stars are a small class of ultracompact binaries, consisting of a white dwarf accreting from another white dwarf or a semi-degenerate helium donor. They are characterised by their hydrogen deficient spectra and short orbital periods ($\sim5-65$ minutes). Currently, 39 AM\,CVn binaries are known (see e.g. \citealt{solheim10} for an overview, and \citet{carter13amcvn,carter14}, \citet{levitan13} and \citet{kupfer13} for recent discoveries).

Our identification spectroscopy revealed three new AM\,CVn binaries, CRTS\,J043517.8+002941, CRTS\,J074419.7+325448 and CRTS\,J084413.6-012807. Their spectra are typical of AM\,CVn systems, with weak, but clearly discernible \hei\, and \heii\, emission lines on a blue continuum, and an absence of any hydrogen lines. CRTS\,J043517.8+002941 was also detected and identified as an AM\,CVn independently by the Palomar Transient Factory, where it is known as PTF1\,J043517.73+002940.7.  \citet{levitan13} obtained Keck-I/LRIS spectra in quiescence and estimated its orbital period to be $34.31\pm1.94$ minutes. 

All three binaries fall within the SDSS footprint where a targeted spectroscopic search for AM\,CVn systems was carried out \citep{roelofs09,rau10,carter13amcvn}, but none of them were detected in that survey. The survey involved spectroscopic identification observations of colour-selected targets in SDSS DR9, down to a dereddened magnitude of $g=20.5$, and is $\sim70$ per cent complete. Two of the three AM\,CVns identified by us are fainter than this limit, and the third, though bright enough, falls outside the colour selection box used to select their targets for follow-up spectroscopy. In fact, the DR9 colours of both CRTS\,J074419.7+325448 and CRTS\,J084413.6-012807 are redder than the $g-r<-0.1$ cut applied (see Table~\ref{tab:amcvns}). This cut is based on the colours of known AM\,CVn binaries, and was made to eliminate (mainly) quasar contaminants and increase the efficiency of the search program. It is however important to note that disc accretors are colour-variable due to the changes in the disc during or near an outburst. Both these targets fall inside the colour cut (i.e. they have `normal' AM\,CVn colours) when their SDSS DR7 photometry is considered. They were fainter and bluer at the time of the DR7 observations than when the DR9 photometry was taken. It appears that they were rising towards or fading after an outburst at the time of their DR9 observations and that the colour-variability was large enough to push them out of the (NUV$-u$, $g-r$) colour selection box.

\begin{table*}
\caption{\label{tab:amcvns} Dereddened magnitudes of the AM\,CVn systems and helium-enriched binaries identified in our follow-up sample.}
\begin{tabular}{lccccccccccc}
\hline
  \multicolumn{1}{c}{CRTS ID} & \multicolumn{2}{c}{\galex} & \multicolumn{3}{c}{SDSS DR7} & MJD & \multicolumn{3}{c}{SDSS DR9} & MJD & \multicolumn{1}{c}{NOTE} \\
   ~                         & FUV    & NUV    & $u$    & $g$    & $(g-r)$  &  [DR7]  & $u$    & $g$    & $(g-r)$  &  [DR9]  & ~ \\
\hline\hline
  CRTS$\:$J043517.8$+$002941 & --     & --     & --     & --     &  --      & --      & 21.502 & 21.684 & $-0.239$ & 51136.3 & $a,d$ \\
  CRTS$\:$J074419.7$+$325448 & 20.349 & 20.568 & 20.923 & 21.131 & $-0.292$ & 51961.1 & 20.479 & 20.602 & $+0.029$ & 54152.3 & $b$   \\
  \smallskip
  CRTS$\:$J084413.6$-$012807 & 19.898 & 20.164 & 20.050 & 20.264 & $-0.160$ & 52016.2 & 20.121 & 20.004 & $-0.034$ & 54057.4 & $b$   \\
  CRTS$\:$J080853.7$+$355053 & 19.608 & 19.533 & 19.550 & 19.612 & $-0.009$ & 52262.3 & --     & --     & --       & --      & $c,e$ \\
  CRTS$\:$J164748.0$+$433845 & 21.900 & 21.550 & 21.326 & 21.551 & $-0.258$ & 51991.4 & --     & --     & --       & --      & $c,e$ \\
\hline
\end{tabular}
\begin{flushleft}
\begin{footnotesize}
$a$ -- AM\,CVn binary, also known as PTF1\,J043517.73+002940.7.
$b$ -- New AM\,CVn binary.
$c$ -- Helium-enriched CV.
$d$ -- Area not covered by \galex\, GR6 or SDSS DR7.
$e$ -- No new DR9 measurement.\\
\end{footnotesize}
\end{flushleft}
\end{table*}

Our follow-up sample also contain two helium-enriched CVs. The Gemini North spectrum of CRTS\,J080853.7$+$355053 displays strong lines of hydrogen, identifying it as a CV rather than an AM\,CVn system. The line flux ratios however reveal that it has an enhanced helium abundance compared to normal CVs. We measure \hei5876/\halpha$=0.938\pm0.003$, compared to $0.2 - 0.4$ for typical dwarf novae \citep[e.g.][]{szkody81,thorstensen01}. There are no features from the companion star visible in the spectrum, but the high He/H ratio suggests that it had undergone significant nuclear evolution before mass transfer started in this binary. 

Two similar systems have recently been discovered, CRTS\,J112253.3$-$111037 \citep{breedt12j1122} and CRTS\,J111126.9+571239 \citep{carter13sbs,littlefield13}. Both these binaries have orbital periods well below the orbital period minimum of CVs and enhanced helium in their optical spectra. As the binary continues to evolve, the spectrum will become increasingly helium rich. \citet{podsiadlowski03} proposed this mechanism as a possible formation channel for AM\,CVn binaries, but because of the lack of observed progenitor binaries, as well as long evolutionary timescales required, this formation channel is considered less likely than alternatives \citep[e.g.][]{nelemans10}. Population synthesis models suggest that less than two per cent of compact binaries could form this way \citep{nelemans04}, favouring instead models which involve two phases of common envelope evolution to reduce the binary to its compact size. However, CRTS\,J080853.7$+$355053 is now the third such system discovered by CRTS. Its orbital period is still unknown, but given the similarity of its spectrum to CRTS\,J112253.3$-$111037 and CRTS\,J111126.9+571239, we expect it to be well below 80 minutes as well. With the addition of the recently discovered BOKS\,45906, a 56~minute CV in the Kepler field \citep{ramsay13}, evolved CVs now account for 10 per cent of all semi-detached white dwarf binaries with orbital periods below the CV period minimum \citep{breedt12j1122}. This fraction increases to 12 per cent if CRTS\,J080853.7$+$355053 is included as well. With such a large fraction of possible progenitors, the evolved CV channel can clearly not be neglected when considering pathways of AM\,CVn formation. 

There are several other AM\,CVn candidates and ultracompact CVs among the full sample of CRTS CV candidates that did not form part of our spectroscopic follow-up. CRTS\,J045019.8-093113 was identified as a possible AM\,CVn by \citet{drake12atel4678}, and photometric follow-up by \citet{woudt12atel4726} revealed a period of $47.28\pm0.01$~min. They note that it is unusual for an AM\,CVn at such a long period to undergo outbursts. \citet{levitan13} also identifies CRTS\,J163239.3+351108 (=PTF1\,J163239.39+351107.3) as a possible AM\,CVn, but the spectrum they obtained was insufficient to confirm this classification. 
An outburst spectrum of CRTS\,J102842.9-081927 by vsnet\footnote{http://www.kusastro.kyoto-u.ac.jp/vsnet/} observers was reported to show both hydrogen and helium lines (Kato \& Kinugasa 2009, vsnet-alert 11166) but photometry carried out by \citet{woudt12} revealed a very short period of 52~minutes. A short photometric period is also seen in CRTS\,J233313.0-155744 \citep[62 min,][]{woudtwarner11_cs2333} but little else is known about this system at present. CRTS also detected outbursts of the known AM\,CVn binaries CP\,Eri, SDSSJ1721+2733 \citep{rau10} and SDSSJ0926+3624 \citep{copperwheat11}\footnote{The corresponding CRTS IDs are CRTS\,J031032.8-094506, CRTS\,J172102.5+273301 and CRTS\,J092638.7+362402.}. So, in the complete sample of 1043 CRTS CVs and CV candidates, seven are confirmed AM\,CVn systems, and a further six have periods below the CV period minimum. 

We also include in this section the spectrum of CRTS\,J164748.0$+$433845. It appears to show only broad \heii\, in emission, on top of an otherwise featureless blue continuum. Closer inspection reveals weak \halpha\, emission as well, but no other lines are apparent in the spectrum. It is possible that the system was in outburst during our observations (the acquisition magnitude is 1.85 mag brighter than its quiescent magnitude from SDSS) but the isolated strong \heii\, line remains unusual. Strong \heii\, emission is often associated with a magnetic white dwarf, but it can also be enhanced by accretion activity during a dwarf nova outburst. The field containing CRTS\,J164748.0$+$433845 has been observed 80 times over the nine years of CSS observations, but it has only been detected on five of those nights, since it is well below the CSS detection limit during its faint state. We note that the SDSS colours of this system (Table~\ref{tab:amcvns}) fall inside the colour selection box for AM\,CVn, as defined by \citet{carter13amcvn}. The nature of this system is currently unclear. Its blue colour and unusually strong \heii\, emission certainly makes it worthy of further observations to determine whether it is another AM\,CVn system.

\subsection{White dwarf dominated spectra} \label{sec:wddominated}
In most of the known non-magnetic cataclysmic variables, the luminosity is dominated by the emission from the accretion disc. If however the accretion rate is low and the accretion disc faint, spectral features from the component stars may be visible in the optical spectra. 

25 of our spectra display broad absorption wings in the Balmer lines, formed in the atmosphere of the white dwarf. These targets are labelled ``W'' in Table~\ref{tab:targetinfo}. In CRTS\,J075414.5+313216, CRTS\,J130030.3+115101, CRTS\,J132103.2+015329, CRTS\,J140454.0-102702 and CRTS\,J172515.5+073249 the flux is strongly dominated by the white dwarf.  Such a strong signature of the white dwarf in the optical spectrum is often seen in old, short period CVs with low accretion rates \citep{gaensicke09}. Photometric observations are available for four of these five CVs, confirming that they have short orbital periods. 
\citet{katoIII12} find a superhump period of 90.86~min in CRTS\,J075414.5+313216. The spectrum of this system also displays broad \caii\, emission, likely from the accretion disc.
\citet{woudt12} found a strong double-humped modulation in the quiescent light curve of CRTS\,J130030.3+115101 and measured an orbital period of 90.24~min. They also report CRTS\,J140454.0-102702 to be an eclipsing CV with a period of 86.05~min.  
CRTS\,J132103.2+015329, also known as HV\,Vir, has an orbital period of 82.18~minutes \citep{patterson03,woudt12}. 

There are also several targets in which the white dwarf signal is slightly weaker but still clearly visible at \hbeta\, and \hgamma. CRTS\,J043546.9+090837, CRTS\,J082821.8+105344 and CRTS\,J224823.7$-$092059 are among the faintest targets we observed spectroscopically. Only \halpha, \hbeta\, and weak \heil5876 emission lines are visible in their spectra, but the lines are strong enough to clearly identify these targets as CVs. 

CRTS\,J010522.2+110253, CRTS\,J084555.0+033929, CRTS\,J102637.0+475426, CRTS\,J104411.4+211307 and CRTS\,J144011.0+494734 are brighter, but display similar weak absorption from the white dwarf plus strong Balmer and \hei\, emission lines. 
Photometric observations during the decline from outburst revealed superhump periods for four of these targets: 86.92~min in CRTS\,J084555.0+033929 (=V498\,Hya) \citep{katoI09}, 119.4~min in CRTS\,J102637.0+475426 \citep{katoIII12}, 85.09~min in CRTS\,J104411.4+211307 and 92.90~min in CRTS\,J144011.0+494734 \citep{katoII10}. The period of CRTS\,J010522.2+110253 is unknown.
                                                                                                                                                                                                                                                                                                         
It is interesting to note that for the majority of targets discussed in this section, only a single outburst has been observed by CRTS. 
Of course, the coarse sampling of CRTS observations means that some outbursts could have been missed, but over a baseline of nine years, observing only a single outburst is strongly suggestive of a long outburst recurrence time, and strengthens the classification of these dwarf novae as WZ\,Sge-type stars. We return to this issue in section~\ref{sec:outbursts}.  

The spectra of five stars display absorption features from both the white dwarf and an M-type donor star: CRTS\,J084127.4+210054, CRTS\,J085113.4+344449, CRTS\,J102616.0+192045, CRTS\,J103317.3+072119 and  CRTS\,J145921.8+354806. Although this implies a low accretion rate in these systems, they are unlikely to be WZ\,Sge stars, as the donor would be too faint to see in the optical spectrum of WZ\,Sge stars. The CRTS light curves of all five these systems show multiple outbursts. Superhump periods reveal that these systems are close to the lower edge of the period gap, so it is possible that the donor is only just filling its Roche lobe after evolving through the gap, and that the accretion rate is still low. 
We note that the dwarf nova IR\,Com ($\Porb=125.4$~min, so also at the lower edge of the period gap) spent more than two years in a `low state', which is very unusual for a non-magnetic CV \citep{manser14}. While earlier spectra of this system are clearly disc-dominated with strong emission lines, the low state spectrum resembles a detached binary with little evidence of accretion. This suggests that the accretion initially occurs in short bursts, and the system takes some time to reach a stable mass transfer rate.   

The CRTS CVs which show both the white dwarf and the donor star in their spectra have superhump periods of 126.3~min in CRTS\,J084127.4+210054 \citep{katoIV13}, $\sim115$~min in CRTS\,J085113.4+344449 (Maehara 2008, vsnet-alert 10723), 119.0~min in CRTS\,J102616.0+192045 \citep{katoIII12} and 122.6~min in CRTS\,J145921.8+354806 \citep{katoIV13}. 
The period of CRTS\,J103317.3+072119 is not known. Its spectrum is very blue, with only a small contribution from the donor star.

\subsection{Donor star visible in the continuum}
The donor star is recognised in the optical spectrum by absorption bands from Mg (near 5200\AA) and TiO (near 7100\AA). 
Five of our targets have spectra which show features consistent with an M star donor: CRTS\,J073339.3+212201, CRTS\,J082123.7+454135, CRTS\,J135219.0+280917, CRTS\,J145502.2+143815 and CRTS\,J160003.7+331114. The latter is also known as VW\,CrB and has a superhump period of 104.9~min \citep{nogami04}.

Additionally, three spectra show evidence of a K star donor: CRTS\,J081936.1+191540, CRTS\,J151020.7+182303 and CRTS\,J163805.4+083758 (=V544\,Her). The latter is a known, but poorly studied dwarf nova. The continuum appears to be that of a hot (K type) donor star, typical of long period systems. However, \citet{howell90} measure a photometric period of only 100~min. A normal main sequence K~star is too large to fit inside the Roche lobe of such a short period binary. A typical CV at this orbital period has a M5--6 donor \citep{knigge11}. It therefore appears that this is a binary in which mass transfer only started after the donor had already started to evolve. The hot donor star seen in the spectrum is the evolved core, stripped from its outer layers by the accretion process. Another such CV has recently been reported by \citet{thorstensen13} (CRTS\,J134052.1+151341) and it is thought that the same process is responsible for some sub-period minimum CVs, such as V485\,Cen \citep{augusteijn93} and EI\,Psc \citep{thorstensen02}. V544\,Her is worthy of further observations.

\subsection{High inclination systems}
The spectra of high inclination (eclipsing) CVs often have broad, double-peaked emission lines, resulting from the velocity difference between opposite parts of the outer accretion disc. 
This spectral feature is often taken as indicative of a high inclination, and therefore a CV that could potentially show eclipses of the white dwarf by the secondary star. Eclipsing systems are very useful, since geometrical modelling of their light curves allows for precise measurements of the system parameters, such as masses and radii of the component stars, binary inclination and orbital period \citep[e.g.][]{littlefair08,savoury11}.

18 of our targets display double-peaked emission lines in their spectra. Three of these have already been confirmed as eclipsers: CRTS\,J043829.1+004016 was found to be a deeply eclipsing CV ($\sim1$ mag) by \citet{coppejans13}, who measured an orbital period of 94.306~min. The same group also reported CRTS\,J082654.7$-$000733 and CRTS\,J140454.0$-$102702 to be eclipsing CVs, with orbital periods of 85.794~min and 86.05~min, respectively \citep{woudt12}. Consistent with the short orbital periods, our spectra of these two CVs show clear absorption from the white dwarf in the Balmer lines (see section \ref{sec:wddominated}), and in the case of  CRTS\,J140454.0$-$102702, the spectrum is strongly dominated by the white dwarf emission. 

CRTS\,J132536.0+210037 is a new eclipsing CV, clearly revealed as such by its SDSS photometry and CRTS light curve in Figure~\ref{fig:spectra}. The SDSS $riz$ magnitudes match our Gemini spectrum well, but the $g$ band measurement is 3.5~mag fainter, as it was taken while the system was in eclipse. The CRTS light curve shows that the quiescent brightness is $\sim19.5$~mag. The spectrum shows clear absorption wings around \hbeta\, and \hgamma, so we expect it to have a short ($<90$~min) orbital period.

The other 14 high inclination systems are flagged in Table~\ref{tab:targetinfo}. Six of these display absorption features from the white dwarf as well. We note that there are reports of a shallow eclipse in outburst photometry of CRTS\,J090239.7+052501 (Uemura \& Arai 2008, vsnet-alert 9955). 
The system is well below the CRTS detection limit during quiescence ($g=23.15$), so high time resolution observations of the eclipse during quiescence, will require a large aperture telescope. Its orbital period, estimated from the superhump period, is 79.9~minutes \citep{katoI09}.

\subsection{Dwarf novae in outburst}

Four of our spectra are typical of dwarf novae in or near outburst. The spectra of CRTS\,J171223.1+362516 and CRTS\,J210043.9$-$005212 display broad Balmer absorption from the optically thick accretion disc, as well as weak \halpha\, emission. They were detected at $g=17.0$ and $18.7$ mag respectively at the time of our Gemini observations, which is 3.9 and 2.9 mag brighter than their SDSS $g$ band magnitudes. 
The spectra of CRTS\,J081414.9+080450 and CRTS\,J210954.1+163052 show a bright continuum, but reasonably strong Balmer emission as well. Their acquisition magnitudes were respectively 4.2 and 1.2 magnitudes brighter than their SDSS $g$ band magnitudes, and match the magnitudes of the outburst detections in their CRTS light curves (Figure~\ref{fig:spectra}). These two CVs were probably on the rise to, or cooling down from, an outburst when their spectra were taken. 
We note that CRTS\,J081414.9+080450 displays strong \heiil4686 in its spectrum, which may be due to the enhanced accretion flow.
We compared our spectroscopy observation dates with the CRTS light curves, but none of the four outbursts were covered by the CRTS observations.

\subsection{Ordinary quiescent dwarf novae}

The combination of Balmer and He emission line spectra and light curves showing at least one outburst event, confirm the classification of the remaining 31 targets in our spectroscopic sample as dwarf novae. They are marked as `DN' in Table~\ref{tab:targetinfo}, or as `SU' if a superhump period has been reported in literature. Our spectroscopic observations were taken during quiescence. Most of the targets are below the CRTS detection limit in quiescence, and were only discovered as a result of a bright outburst. 

For completeness, we note that CRTS\,J090516.0+120451 and CRTS\,J113950.6+455818 were observed by \citet{thorstensenskinner12} as well, but we include our spectra here as they extend further into the blue. CRTS\,J115330.2+315836 was observed by SDSS six weeks after our Gemini observation (see section~\ref{sec:boss}), and CRTS\,J171202.9+275411 was also observed by \citet{wils10}.


%
\section{Other new spectroscopic identifications} \label{sec:otherspec}

In addition to the 72 spectra from our own spectroscopic follow-up program presented in the previous section, a further 13 targets from the full sample of 1043 CRTS CVs have previously unpublished spectra available in public spectroscopic surveys.

\subsection{SDSS BOSS} \label{sec:boss}
The Baryonic Oscillations Spectroscopic Survey (BOSS) is the most recent (and ongoing) optical spectroscopic survey from the SDSS. The first three years of BOSS observations were published on 2013 July 31 as part of SDSS Data Release 10 \citep[DR10,][]{sdssdr10}. Among the $\sim145\,000$ BOSS stellar spectra available, we found eight matches to our CRTS CV sample. For completeness, we include the BOSS spectrum of CRTS\,J115330.2+315836, which was taken six weeks after our Gemini observation, shown in Figure~\ref{fig:spectra}. Of the eight CVs identified among the BOSS spectra, only CRTS\,J221519.8$-$003257  (=V344\,Aql) is a previously known CV. 

\begin{figure*}
\centering
\rotatebox{270}{\includegraphics[height=17.5cm]{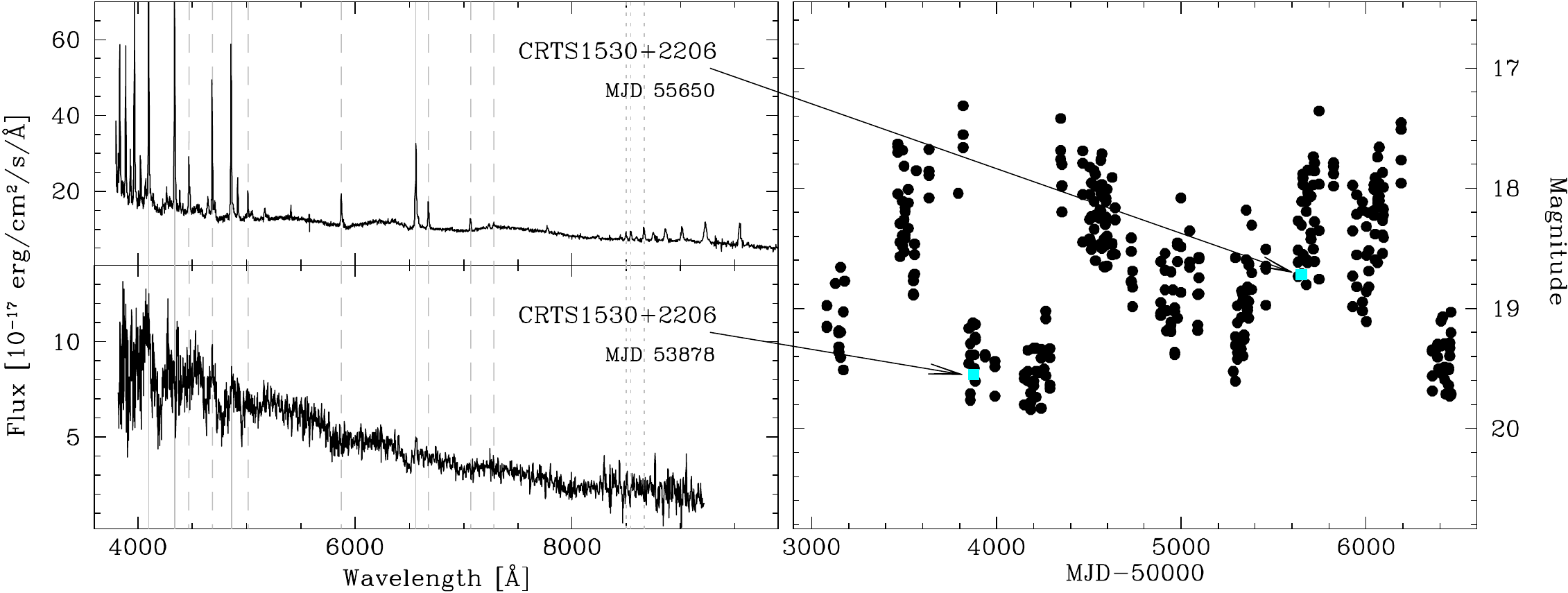}}
\caption{\label{fig:J1530} We identified CRTS\,J153023.6+220646 as a polar from the cyclotron humps and strong \heiil4684 emission in its SDSS BOSS spectrum (top panel). Another spectrum is available in the SDSS database (bottom panel), taken while the system was in a low accretion state. The coloured squares and arrows indicate where the SDSS measurements fit into the CRTS light curve. The variations seen in the CRTS light curve are typical of polars switching between high and low accretion states. }
\end{figure*}

CRTS\,J153023.6+220646 has an older SDSS spectrum available in addition to the BOSS spectrum (Figure~\ref{fig:J1530}), but it was not previously recognised as a CV.  The BOSS spectrum displays obvious cyclotron humps in the continuum as well as a very strong \heiil4684 emission line --- hallmarks of a magnetic CV. The CRTS light curve is typical of polar variability. It shows variations of $\sim3$~mag as the systems switches between high and low accretion states. The older SDSS spectrum was taken during the low state, when the accretion was switched off. The system was much fainter, and Zeeman splitting from the magnetic white dwarf is visible in Balmer lines of the spectrum. 

\begin{figure*}
\centering
\rotatebox{270}{\includegraphics[height=17.5cm]{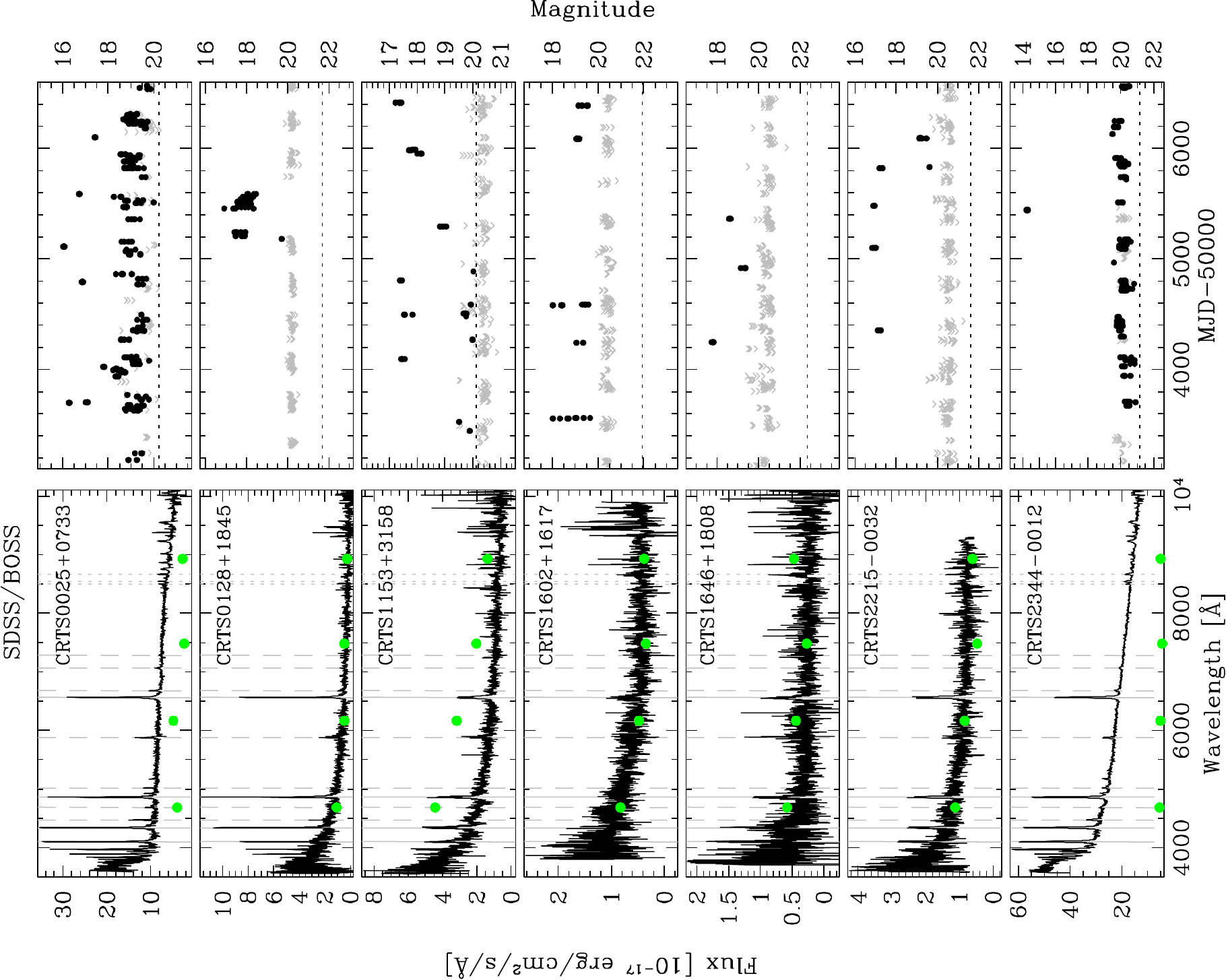}}
\caption{\label{fig:boss} Seven of the eight CRTS CVs which were observed as part of the BOSS survey. (The eighth is shown in Figure~\ref{fig:J1530}.) The spectra, released as part of SDSS DR10, were smoothed by a 3-point boxcar for display purposes. The CRTS light curve of each target is shown in the panel to the right. Symbols and lines are as defined in Figure~\ref{fig:spectra}.}
\end{figure*}

The other seven spectra are shown in Figure~\ref{fig:boss} along with their CRTS light curves. The spectra display clear Balmer emission and the variability of the light curves identify them as normal dwarf novae. Table~\ref{tab:boss} list the basic spectral properties of the BOSS CVs, as well as the plate-MJD-fibre information to locate their DR10 spectra.

\begin{table*}
\caption{\label{tab:boss} CRTS CVs with SDSS BOSS spectra.}
\begin{tabular}{cclccrc}
\hline
  \multicolumn{1}{c}{SDSS ID} &
  \multicolumn{1}{c}{CRTS ID} &
  \multicolumn{1}{c}{$g$ mag} &
  \multicolumn{1}{c}{plate} &
  \multicolumn{1}{c}{MJD} &
  \multicolumn{1}{c}{fiber} &
  \multicolumn{1}{c}{COMMENTS} \\
\hline\hline
  SDSS\,J002500.18$+$073349.3 & CRTS\,J002500.2$+$073350 & $20.23$  & 4539 & 55865 & 533  &  Dwarf nova\\
  SDSS\,J012838.37$+$184535.6 & CRTS\,J012838.3$+$184536 & $21.67$  & 5135 & 55862 & 59   &  Dwarf nova\\
  SDSS\,J115330.25$+$315835.9 & CRTS\,J115330.2$+$315836 & $20.12$  & 4614 & 55604 & 44   &  Dwarf nova\\
  SDSS\,J153023.92$+$220642.5 & CRTS\,J153023.6$+$220646 & $18.72^*$& 3949 & 55650 & 814  &  Polar (high state)\\
            $-$         &                $-$       & $19.55^*$& 2161 & 53878 & 34$^\dagger$ & Polar (low state)\\
  SDSS\,J160232.13$+$161731.7 & CRTS\,J160232.2$+$161732 & $21.94$  & 3922 & 55333 & 898  &  Dwarf nova\\
  SDSS\,J164624.78$+$180808.9 & CRTS\,J164624.8$+$180808 & $22.33$  & 4062 & 55383 & 20   &  Dwarf nova\\
  SDSS\,J221519.80$-$003257.1 & CRTS\,J221519.8$-$003257 & $21.61$  & 4200 & 55499 & 122   &  V344\,Aql. High inclination \\
  SDSS\,J234440.53$-$001205.8 & CRTS\,J234440.5$-$001206 & $21.14$  & 4213 & 55449 & 38   &  Dwarf nova\\
\hline
\end{tabular}
\begin{flushleft}
$^*$  Estimated from spectrum, not SDSS photometry\\
$^\dagger$  Spectrum taken as part of the SDSS Legacy survey, using the original SDSS spectrograph\\
\end{flushleft}
\end{table*}

\subsection{PESSTO} \label{sec:pessto}
PESSTO is the Public ESO Spectroscopic Survey for Transient Objects\footnote{http://www.pessto.org} carried out by a consortium of institutions on the European Southern Observatory's (ESO) New Technology Telescope on La Silla, Chile, using the EFOSC2 (optical) and SOFI (near-infrared) spectrographs. 
The main aim of the survey is to follow-up and classify supernovae across a range of parameters such as luminosity and metallicity. Transients are selected from various surveys such as La Silla-Quest, SkyMapper, Pan-STARRS1 as well as CRTS. As a public survey, the spectra are made available immediately in the ESO archive and are also uploaded to the WISeREP database \citep[Weizmann Interactive Supernova data REPository;][]{wiserep}\footnote{\label{fn:wiserep}http://www.weizmann.ac.il/astrophysics/wiserep/}.

\begin{figure}
\centering
\rotatebox{0}{\includegraphics[width=8.0cm]{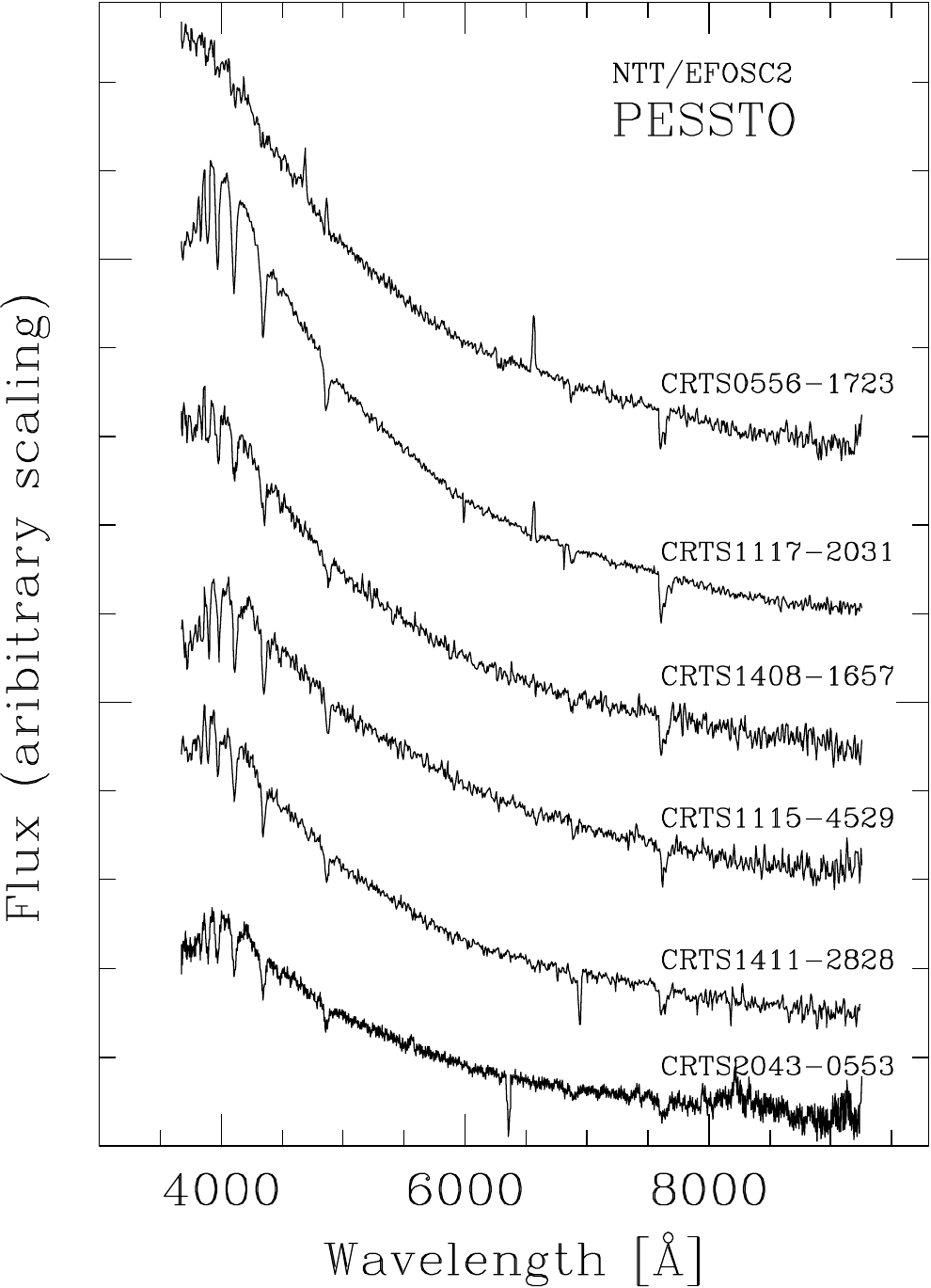}}
\caption{\label{fig:pessto} CVs in outburst observed as part of the PESSTO transient follow-up program. The spectra were smoothed with a 2-pixel boxcar for display purposes.}
\end{figure}

Most targets observed by PESSTO are supernovae, but occasionally a large amplitude transient turns out to be a dwarf nova in outburst. We found spectra for six CRTS CVs in the WISeREP database, four of which were reported as ``stellar'' in Astronomer's Telegrams\footnote{ATEL, http://www.astronomerstelegram.org/} (see Table~\ref{tab:pessto}).
All six targets have steep blue spectra with Balmer absorption, typical of CVs in outburst. The spectra of CRTS\,J055628.1$-$172343 and CRTS\,J111742.1$-$203116 show  \halpha\, in emission. The spectra are shown in Figure~\ref{fig:pessto} and information on the observations are given in Table~\ref{tab:pessto}. 
All six CVs are below the CRTS detection limit in quiescence and have only been observed during these discovery outbursts.

\begin{table}
\caption{\label{tab:pessto} Outburst spectra of CVs as observed by PESSTO as part of supernova follow-up observations. All spectra were taken with the EFOSC2 spectrograph on the NTT.}
\begin{tabular}{|r|c|l|}
\hline
  \multicolumn{1}{|c|}{CRTS ID} &
  \multicolumn{1}{c|}{Obs. date } &
  \multicolumn{1}{c|}{Ref. note$^*$} \\
\hline
  CRTS\,J055628.1$-$172343 & 2013-03-06 & WISeREP:\,CV  \\
  CRTS\,J111518.0$-$452917 & 2013-03-13 & ATEL \#4882 \\
  CRTS\,J111742.1$-$203116 & 2013-04-03 & ATEL \#4940 \\ 
  CRTS\,J140859.4$-$165709 & 2013-03-17 & ATEL \#4892 \\
  CRTS\,J141135.5$-$282828 & 2013-04-04 & ATEL \#4952 \\
  CRTS\,J204335.9$-$055316 & 2012-10-09 & WISeREP:\,CV  \\
\hline
\end{tabular}
\begin{footnotesize}                                                                                                      
\begin{flushleft}
$^*$ WISeREP: see footnote~\ref{fn:wiserep},
     ATEL \#4882: \citet{atel4882},
     ATEL \#4940: \citet{atel4940},
     ATEL \#4892: \citet{atel4892},
     ATEL \#4952: \citet{atel4952}.
\end{flushleft}
\end{footnotesize}
\end{table}


%
\section{CV light curves from CRTS} \label{sec:crtslcs}

Three telescopes, equipped with identical 4k$\times$4k pixel CCD cameras, contribute imaging data to the Catalina Sky Survey. 
CRTS started searching the imaging data of the Catalina Schmidt Survey (CSS) for transients on 2007 November 8. The survey is carried out using the 0.7m Catalina Schmidt telescope on Mount Bigelow, just north of Tucson, Arizona, USA. On the nearby Mount Lemmon, the 1.5m telescope is used to scan the ecliptic latitudes for near-earth objects and transients. The Mount Lemmon Survey (MLS) data was added to the transient processing pipeline on 2009 November 6. The Siding Springs Survey (SSS) was carried out from the Siding Springs Observatory in Australia, using the 0.5m Uppsala Schmidt telescope. It was added to the CRTS project on 2010 May 5 and completed operations in 2013 July. Further details of the telescopes, as well as the transient selection criteria and CV identification process are discussed in detail by \citet{drake09crts} and \citet{drake14}.

The images are processed using an aperture photometry pipeline based on {\sc SExtractor} \citep{sextractor} and the unfiltered magnitudes are transformed to Landolt $V$ using observations of standard stars \citep{drake13calib}. Information on the transients detected, e.g. finder charts and historical light curves, are released immediately on the CRTS webpage\footnote{http://nesssi.cacr.caltech.edu/catalina/Allns.html} and on VOEventNet\footnote{http://voeventnet.caltech.edu/feeds/Catalina.shtml}. Additionally, the light curves of 500 million objects were made available as a public data release (CSDR2) in  2012\footnote{http://nesssi.cacr.caltech.edu/DataRelease/}. There is some overlap in the sky footprint of the three surveys, so where available, CSDR2 also includes photometry extracted from the images of the other sub-surveys, as well as images taken before the start of CRTS. 

\begin{figure}
\centering
\rotatebox{0}{\includegraphics[width=5.5cm]{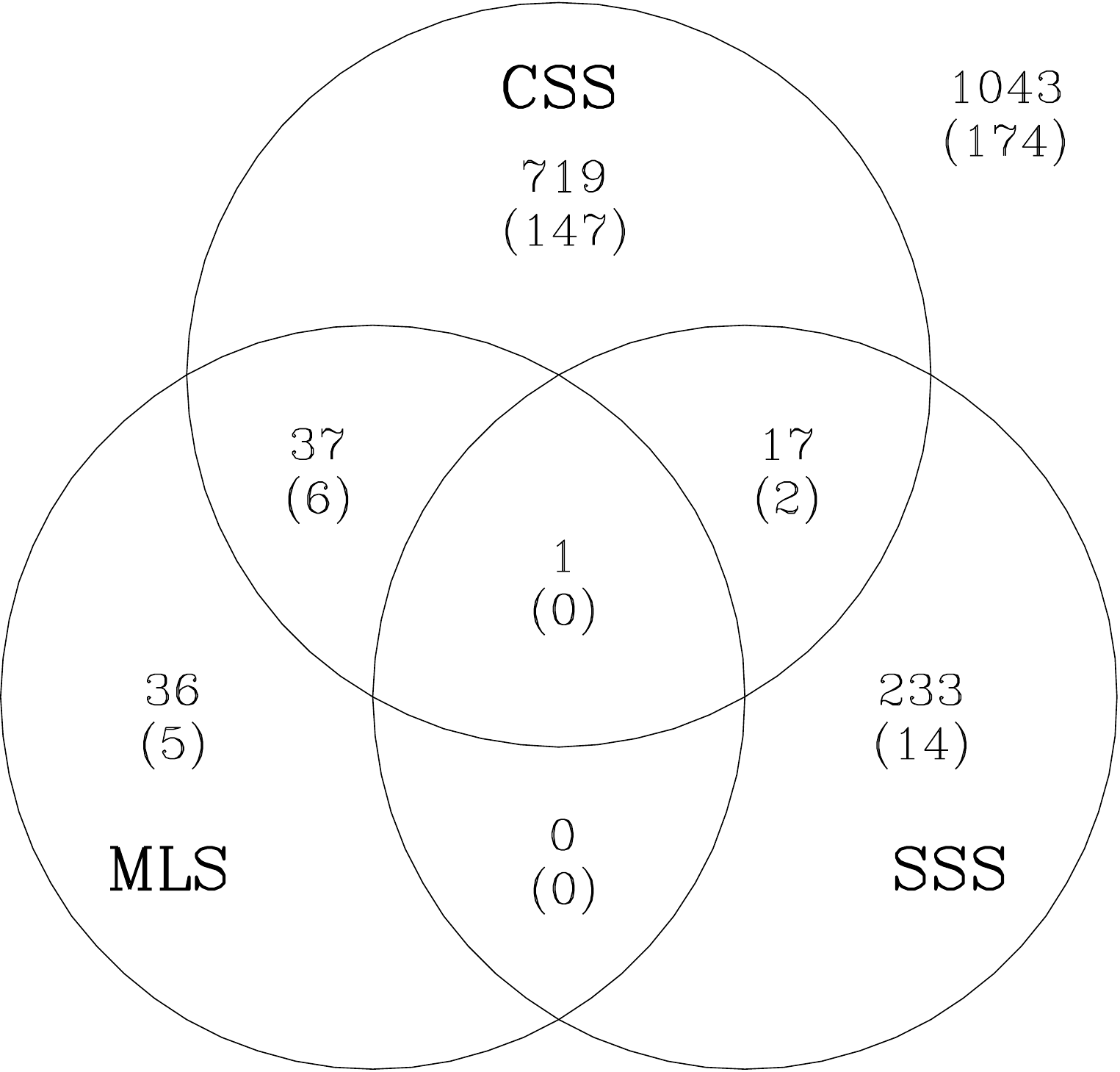}}
\caption{\label{fig:venn} A total of 1043 CV candidates have been identified in the six years since the start of CRTS. This Venn diagram illustrates the overlap between the samples identified by the three different telescopes contributing to the survey. Only one CV so far appears in all three surveys. The number in brackets is the number of CVs which were known prior the start of CRTS, but re-identified as a result of their variability.}
\end{figure}

For the analysis in this paper, we extracted the light curves of all targets identified as CV candidates, a total of 1098 transient identifications. Many of the targets have been detected as transients multiple times (e.g. recurring outbursts) and some targets have been detected in more than one of the sub-surveys. In all cases, we use the earliest transient detection as the discovery date and the discovery survey of the object. There are 37 CVs which were independently identified by CSS and MLS, and 17 which were identified by both CSS and SSS. Only one CV so far appears in all three surveys. The combined data from the three surveys gives a catalogue of 1043 unique objects. The overlap between the identifications is illustrated with a Venn diagram in Figure~\ref{fig:venn}.

The number of points per light curve varies considerably, as it depends not only on how frequently the field was observed, but also on the quiescent brightness of the target. At the moment, only real detections are included in the CSDR2 light curves, but future data releases will include upper limits for the non-detections as well. The upper limits are particularly useful for targets which can drop below the detection limit of the telescope as a result of their intrinsic variability, or very faint targets which are only detected in clear, good seeing conditions. In order to illustrate the typical observing cadence, we extracted cut-out images centred on each of our spectroscopic targets and carried out differential photometry as described in \citet{parsons13crts}. The photometric measurements are shown as black dots in Figure~\ref{fig:spectra}. For the nights on which the target was observed, but not detected, we plot an upper limit (grey arrow) derived from nearby stars in the image. There have been between 80 and 590 observations per target over the $9$ years of the Catalina Sky Survey, with a typical observing cadence of 10--14 days. 

Stars closer than 2--3\arcsec on the sky are not resolved by the CRTS telescopes. A `blending' flag included in the CRDR2 light curves alert the user of this possibility, but we decided to inspect the individual light curves by eye, as this limit is also affected by the seeing quality. We plotted the coordinates associated with each photometric point in a given CSDR2 light curve, and compared it to the SDSS image of the same area, where available. The SDSS images have a higher resolution, and the multicolour photometry usually clearly identifies the CV as the bluest object of the pair. In many cases it was also possible to separate the two stars in coordinate space and remove the photometric points belonging to the nearby star from the light curve. For a small number of targets, the nearby star is too close and the stars cannot resolved by the CRTS telescopes. At least four light curves shown in Figure~\ref{fig:spectra} are affected by blending, CRTS\,J112509.7+231036, CRTS\,J144011.0+494734, CRTS\,J145921.8+354806 and CRTS\,J171223.1+362516. We added the word `blend' to their light curve panels to indicate that another star affect the appearance of the light curve.


%
\section{Outburst properties of the CRTS sample of CVs} \label{sec:crtscvs}

So far, after six years of operations, CRTS has identified a total of 1043 CVs and CV candidates
--- the largest CV sample from any survey to date. Only 174 (17 per cent) of these CVs were known prior to the start of CRTS, 869 are new discoveries.  
The results from various follow-up programs \citep[section~\ref{sec:targetnotes} of this paper,][]{katoI09,katoII10,katoIII12,katoIV13,woudt12,thorstensenskinner12,coppejans13} show that the identification process is highly accurate. Between 97 and 100 per cent of individual samples of CV candidates which were followed up, were confirmed as CVs. 

Of the 869 new discoveries, at least 154 have been confirmed spectroscopically so far. A further 86 of the new CRTS discoveries have been confirmed by photometric follow-up. In total (including the previously known CVs), at least 40 per cent of candidates have been spectroscopically or photometrically confirmed as CVs. In the sections that follow, we will assume that all objects in the catalogue are genuine cataclysmic variables.

\subsection{Discovery rate}

\begin{figure}
\centering
\rotatebox{270}{\includegraphics[width=11.0cm]{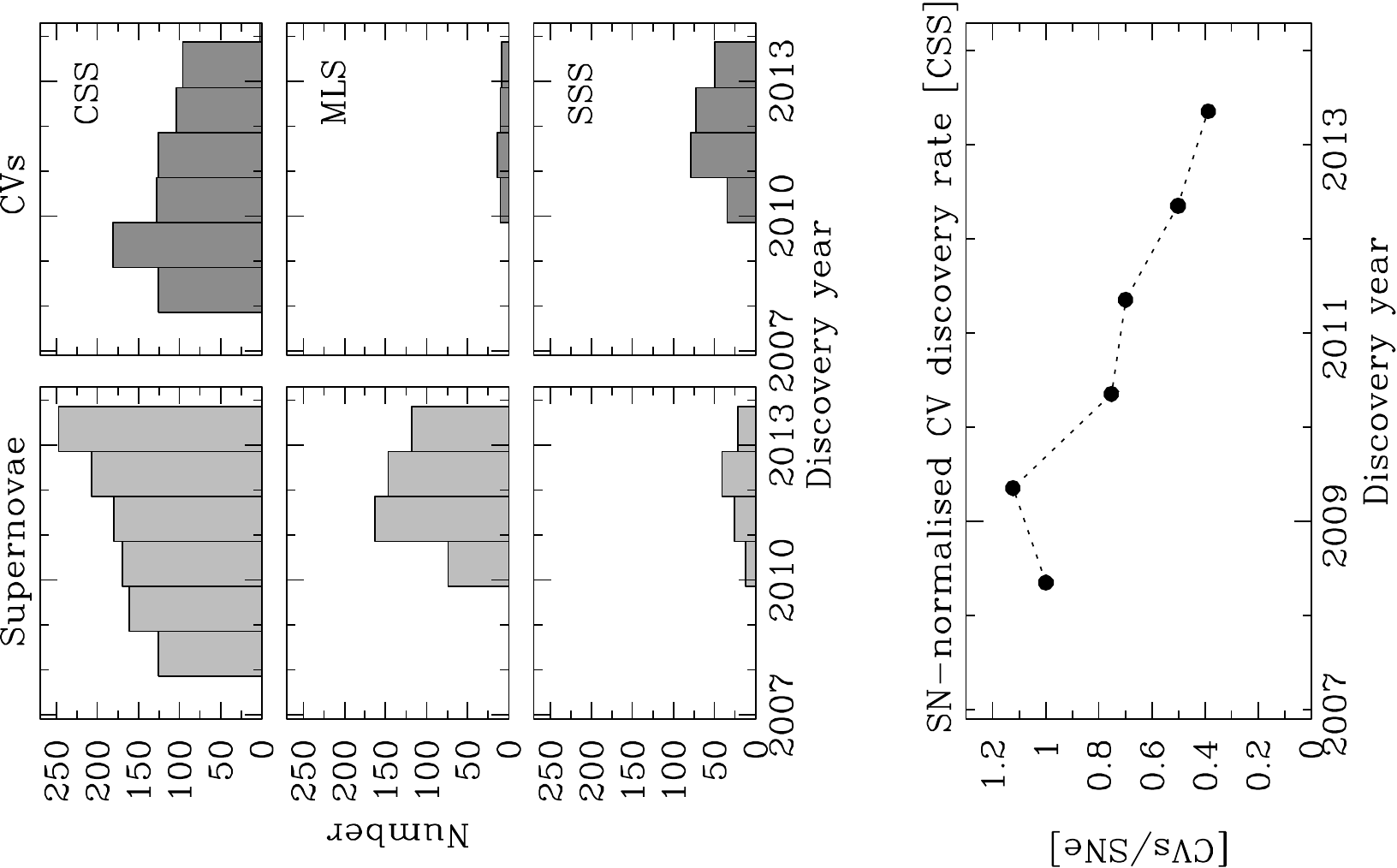}}
\caption{\label{fig:discovery} CRTS discovery rate of supernovae (left) and CVs (right), for each of the three contributing surveys as labelled. Using the number of supernovae as a proxy for the detection efficiency, the bottom panel shows that the effective CV discovery rate is decreasing.} 
\end{figure}

The largest fraction of transients detected by CRTS are supernovae. They account for just over 20 per cent of the total number of CRTS transients, compared to the $\sim13$ per cent which are identified as CVs. Supernovae are observed as a single brightening events (as opposed to repeated outbursts observed in many dwarf novae) and are often associated with a faint galaxy in the survey images. Supernovae also stay bright for longer than the dwarf novae, which helps to differentiate between the transient types. There are several observing programs aimed at confirming these supernovae spectroscopically (e.g. PESSTO, see section~\ref{sec:pessto}). If we assume that supernovae occur at a constant rate, the number of supernovae discovered can act as a proxy for the overall transient detection efficiency of CRTS, and allow us to calculate an effective CV discovery rate. Since the CVs and supernovae are identified using the same process, and from the same survey data, this method is particularly useful for removing effects such as the slowly expanding survey footprint and varying weather conditions on the number of transients observed.

The top panels of Figure~\ref{fig:discovery} show histograms of the number of supernovae and CVs identified by each of the three contributing surveys. We bin the histograms to reflect the number of transients per year, since the start of CRTS operations (2007 November). The MLS and SSS data are binned to match the CSS histogram bins for direct comparison. Binning per year also has the advantage that all bins are affected by seasonal weather effects equally. 

Supernovae are typically detected at fainter magnitudes than CVs \citep[see fig.\,3 of][]{drake14}, so the larger MLS telescope discovers relatively more supernovae compared to CVs than the other two surveys. In fact, as shown in Figure~\ref{fig:discovery}, MLS contributes very little to the overall CV count, only $\sim10$ CVs/year. The opposite is true of the SSS, which used the smallest telescope -- this survey discovered more than double the number of CVs compared to supernovae each year.

Since MLS contributes very few CVs overall and SSS has finished operations, the future discovery rate of CVs will be almost entirely determined by CSS. The CSS-discovered CVs are shown in the top panel of Figure~\ref{fig:discovery}. They are clearly fewer in number than the supernovae, and their discovery rate has decreased from $\sim130$ to $\sim100$ CVs/year over the lifetime of CRTS. In contrast, the number of supernovae increases year-on-year, corresponding to the increase in the sky area surveyed by CSS. Normalising the number of CV identifications by the supernova identifications (bottom panel of Figure~\ref{fig:discovery}), shows a steep decline in the effective discovery rate of CVs. Although we expect this decline to continue, we do not expect it to decrease rapidly to zero as the current rate suggests. In the next section we show that CRTS contains a large fraction of low accretion rate systems that will dominate CRTS CV discoveries in future years. Due to their long outburst intervals, their discovery rate is expected to be low, even if they exist in the large numbers suggested by population models. Therefore, we expect the discovery rate to flatten off into a long `tail', with many more to be discovered.

\subsection{Outburst frequencies} \label{sec:outbursts}

The outburst frequency is a key property of the CRTS sample of CVs, as it determines the likelihood of discovering a particular system. On a physical level, the outburst frequency depends mainly on the mass transfer rate from the donor star and the size of the accretion disc \citep{shafter86a,shafter92}, so it is important to consider the selection effects this may have on the overall sample. 

The sky coverage of CRTS is very inhomogeneous, with 40 and 600 epochs at each location. The lightcurves in Figures~\ref{fig:boss} and \ref{fig:spectra} clearly illustrate how this translates to very different sampling patterns for different targets. The number and frequency of observations vary considerably between different light curves. 

Clearly, large gaps in the light curves will bias the sample against CVs with less frequent outbursts. However, these low accretion rate CVs typically have larger outburst amplitudes \citep[e.g.][p.144]{warnerbook}, which means that they exceed the 2 magnitude threshold by which CRTS CVs are selected, for longer. This makes them easier to find compared to those with more frequent, but smaller outburst amplitudes, and biases the sample in the opposite way. In fact, \citet{thorstensenskinner12} find that there is a significant bias against dwarf novae with outburst amplitudes $\Delta m<6$.

It is, in principle, possible to carry out simulations to estimate the number of outbursts that may have been missed when they occurred between observations. However, such simulations include a large number of uncertain and variable parameters, such as the average outburst frequency, outburst duration, outburst amplitude, and the particular sampling pattern for the target under consideration. We therefore decided to characterise the sample using only the observed outburst properties.

\begin{figure}
\centering
\rotatebox{0}{\includegraphics[width=8.5cm]{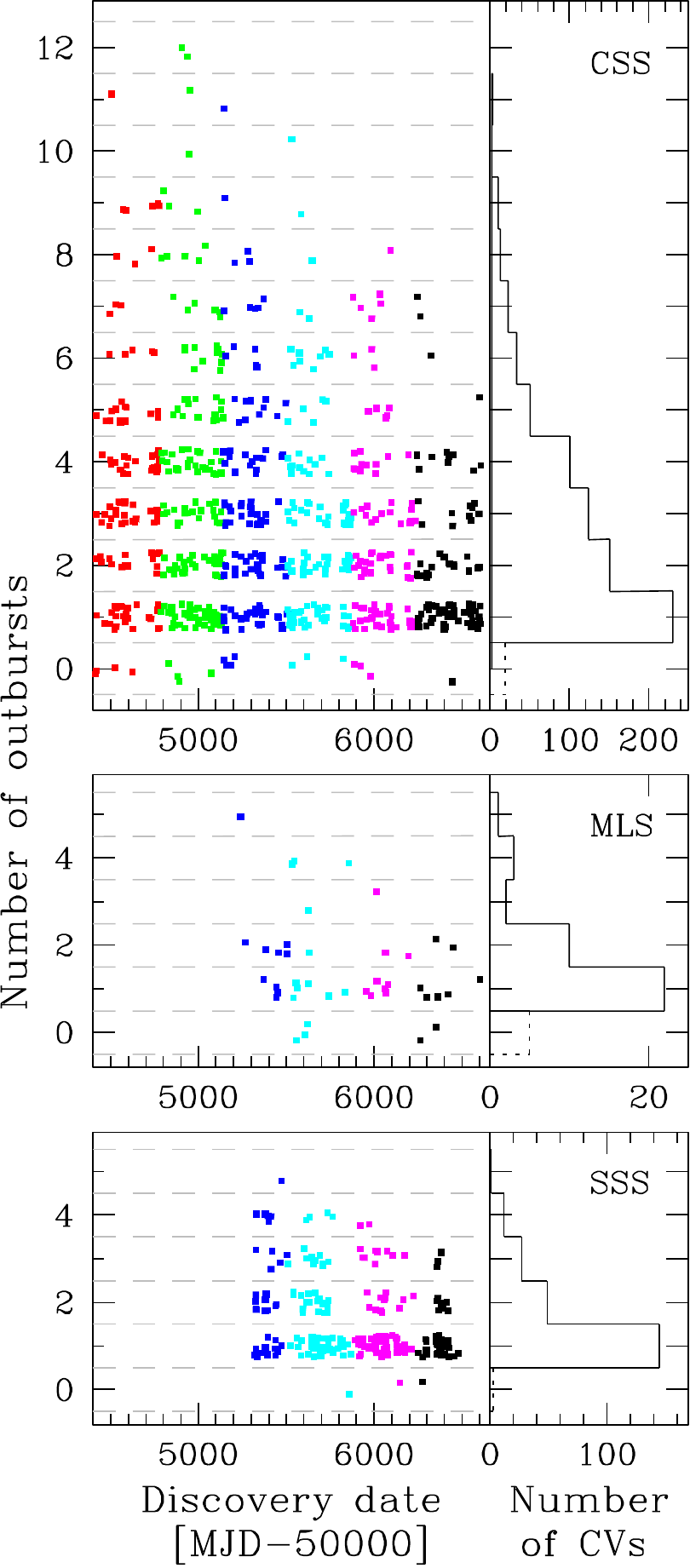}}
\caption{\label{fig:outbursthisto} Number of outbursts detected in the light curves of each of the 1043 CVs, split by the three surveys so that the different starting dates are evident. The scatter plots on the left indicate the number of outbursts as a function of discovery date, and the right hand panels show histograms of the number of CVs for which that number of outbursts have been observed. We applied a small vertical scatter to the outburst number in order to illustrate the density of points in each bin more clearly. All points within a pair of horizontal dashed lines have the number of outbursts indicated by the integer number on the vertical axis. Polars, which show high/low state transitions but no outbursts, are plotted as zero on the vertical scale. }
\end{figure}

Although some outbursts of a given target may have been missed between observations, the number of outbursts observed over the length of the CRTS project gives an indication of the current mass transfer rate in that binary. Figure~\ref{fig:outbursthisto} shows the number of outbursts observed in the light curve of each of the 1043 CRTS CVs, as a function of the discovery date of the system\footnote{We excluded outbursts which occurred before the CRTS started processing the Catalina images for transients. While these earlier outbursts are useful to help constrain the average accretion rate, they did not contribute to the discovery of the CV, so are not included in Fig.~\ref{fig:outbursthisto}.}. Points in the `zero outbursts' bin are known polars, or systems which show polar-like high and low states in their light curves rather than dwarf nova outbursts. For clarity, each year of CRTS operations is shown in a different colour and the points in each bin are offset vertically by a small random amount to better illustrate the density of systems. The histograms in the right hand panels represent the number of CVs in each outburst bin. 

In all three surveys, the shape of the discovery outburst distribution slopes down towards more recent discovery dates, i.e. systems displaying more outbursts in total in their light curves were discovered earlier in the survey. This is unsurprising, since frequently outbursting dwarf novae will have had many outbursts detected, even if gaps in the relatively sparse sampling of CRTS implies that some of their outbursts may have been missed. Similarly, the discovery of a new CRTS CV implies that it had not been detected as a transient (i.e. in outburst) before, so a smaller number of outbursts are expected in more recently discovered systems. Note however, that `quiescent' photometry of these recently discovered targets are available from the Catalina Sky Survey from before the start of CRTS --- the discovery date is simply the date the target was first detected as a transient.

Figure~\ref{fig:outbursthisto} shows that CVs which have only had one observed outburst, span all six years of CRTS operations (i.e. all six different colours appear in the horizontal bar $n=1$ where $n$ is the number of outbursts.) The histograms of all three surveys are dominated by systems with just a single outburst observed by CRTS. This strongly suggests that the sample contains a large fraction of CVs with a long outburst recurrence time, i.e. low accretion rate WZ\,Sge-type systems.

\begin{figure}
\centering
\rotatebox{270}{\includegraphics[width=6.0cm]{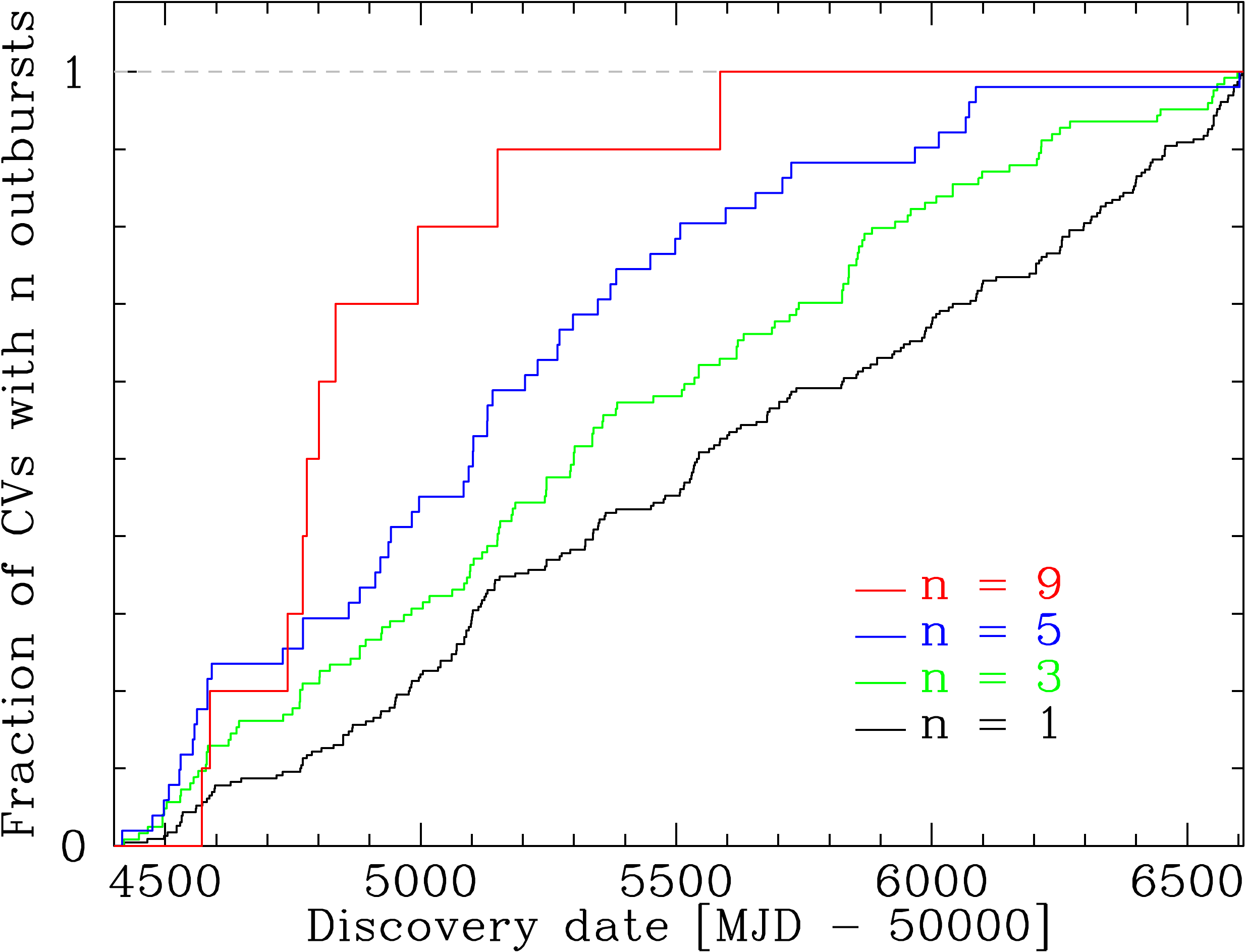}}
\caption{\label{fig:cumulative_outb} Cumulative distribution of CSS CVs with one (black), three (green), five (blue) and nine (red) outbursts observed by CRTS. The number of CVs with a single observed outburst continues to increase at a steady rate. Dwarf novae with high accretion activity, and hence a large number of outbursts, were discovered early in the survey (steep initial rise). Their distributions also flatten off early, suggesting that most dwarf novae with an accretion rate high enough to produce one outburst per year have already been found. }
\end{figure}

To illustrate this point further, we separate the CVs according to the number of outbursts in their light curves and plot the number of systems in each group as a cumulative distribution against their discovery dates (Figure~\ref{fig:cumulative_outb}). 
We only show data from CSS here, in order to use the longest available light curves and to remove any biases that result from the later starting dates of the MLS and SSS surveys.
The cumulative distribution of systems with just one observed outburst increases at a steady rate, apart from the seasonal observing gaps visible as short horizontal stretches in the distribution. CRTS is still discovering $\sim150$ new CVs per year (Figure~\ref{fig:discovery}), and this increase in single-outburst CVs is likely to continue for another few years before most dwarf novae are found and the slope will flatten off. In contrast, CVs with a large number of outbursts, like the $n=9$ case shown in Figure~\ref{fig:cumulative_outb}, rise steeply in the early part of the survey, but also flattens off early. Due to their high levels of accretion activity, most of these systems have been discovered and future observations will simply see more outbursts from the same systems, rather than finding many new CVs with similarly high accretion rates. The cumulative distribution suggest that the majority of dwarf novae in the CRTS survey area, with an accretion rate high enough to produce an average of one outburst per year, have already been found. Most future CRTS discoveries of dwarf novae will be systems with a lower accretion rate.

\begin{figure}
\centering
\rotatebox{270}{\includegraphics[width=6.0cm]{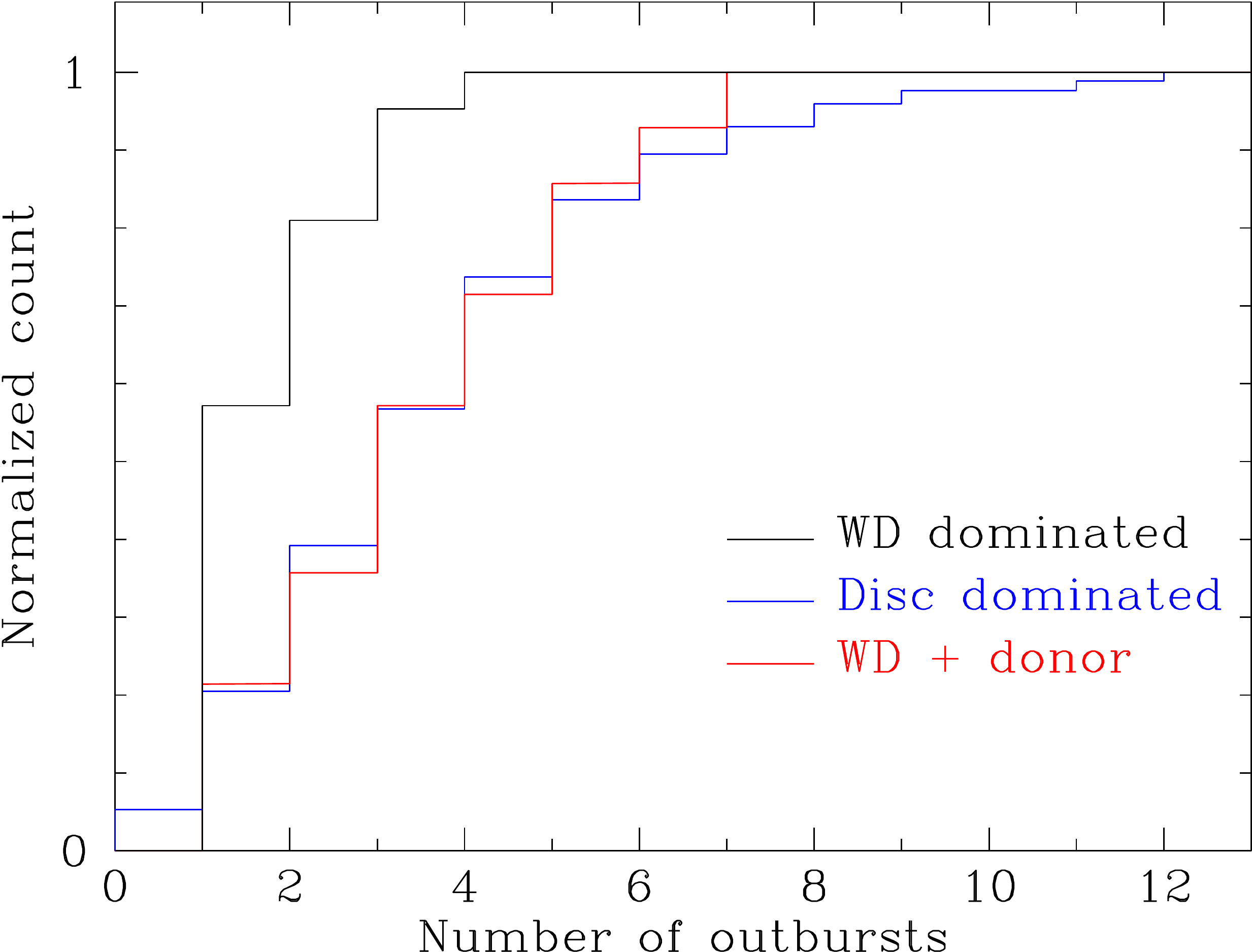}}
\caption{\label{fig:cumulativewd} A comparison between the number of outbursts observed from CVs whose spectra reveal the white dwarf (black), both the white dwarf and the donor star (red) and those where the white dwarf is not seen in the spectrum (blue). A two-sided Kolmogorov-Smirnov test shows that the likelihood that the white dwarf dominated sample was drawn from the same parent population as the disc-dominated sample is only 0.19 per cent, while for those showing both the white dwarf and the donor the likelihood is much higher, 56 per cent.}
\end{figure}

Finally in this section, we compare the outburst properties of the white dwarf dominated CVs with the rest of the sample. As discussed in Section~\ref{sec:wddominated}, CVs which show the absorption wings of the white dwarf in their spectra, must have a low accretion rate, since for higher accretion rates the bright accretion disc dominates the optical luminosity so that the white dwarf is not visible in the spectrum. Given our interpretation of the number of outbursts as the average accretion rate, do the white dwarf dominated CVs also show fewer outbursts? 
We visually inspected the spectra of all CRTS CVs for which we are aware that a spectrum is available, including the CVs known prior to the start of CRTS. Out of a total 250 CRTS CVs with spectra, we found 21 to be white dwarf dominated and an additional 14 to display features from both the white dwarf and the donor star. For 21 CVs we know only of spectra taken during outburst, so these were disregarded. The cumulative distributions of the number of outbursts in the light curves of the three groups of CVs are shown in Figure~\ref{fig:cumulativewd}. 

57 per cent of the white dwarf dominated CVs (12/21) show only a single outburst in their CRTS light curves, in agreement with our interpretation of white dwarf dominated CVs as low accretion rate systems. The lightcurve of one of the white dwarf dominated CVs displays four outbursts and a further three have had three outbursts each, which implies a much higher accretion rate. It is possible that these are low inclination systems, in which the disc is seen almost face-on and the white dwarf is not obscured. There are no white dwarf dominated CVs which have had more than four observed outbursts over the duration of the CRTS observations.

The outburst distributions of the other two samples are very different. Only 20 per cent have only one observed outburst, while 40 per cent of each sample have had more than four outbursts. 
The distribution of CVs in which both component stars are visible in the optical spectrum is statistically indistinguishable from the disc-dominated population in terms of their outburst behaviour. A two-sided Kolmogorov-Smirnov test reveals that the likelihood of the two samples originating from the same parent population is 56 per cent. The white dwarf dominated sample is significantly different. The probability that it is drawn from the same population as the disc-dominated systems is only 0.18 per cent. This supports the interpretation that the CVs which display the white dwarf in their optical spectra are a low accretion rate, intrinsically faint population which display infrequent outbursts \citep{wils10}. In addition, \citet{woudt12} found that the CVs with just a single outburst tend to have short orbital periods, giving further support to the notion that this is an old, evolved population with little accretion activity and low mass donor stars \citep{gaensicke09}.


%
\section{CRTS completeness} \label{sec:completeness}

The observing cadence and sky coverage of the Catalina Sky Survey are optimised for its primary science objective, i.e. detecting near-Earth objects. CRTS is an added-value project which uses the Catalina images to identify transient events and making them publicly available for follow-up observations (Section~\ref{sec:crtslcs}). It is a very efficient `discovery machine', but the inhomogeneities in the survey (e.g. sky coverage pattern, different depths of the three contributing surveys) make it very difficult to reliably determine space densities of the transient populations. CRTS is responsible for most of the currently-known CVs with $g\geq20$ (see section~\ref{sec:alllegacy} below), so beyond this magnitude, even a completeness estimate is a difficult pursuit. However, given the large number of new, faint CVs found by CRTS, and to aid in the planning of future transient surveys, it is still a worthwhile effort to investigate in detail the types of CVs discovered by CRTS compared to other surveys, and to identify any biases that may exist. 

In order to compare homogeneous, representative samples from existing catalogues of CVs, we select only the CVs that fall within a well-defined area of sky --- chosen to be the 7646 square degree footprint of the SDSS Legacy survey, as described below (Section~\ref{sec:legacy}). Throughout Section~\ref{sec:completeness}, we consider only CVs from CRTS, SDSS and the Ritter-Kolb catalogue (Section~\ref{sec:rk}) that fall within this footprint. First, we briefly describe the properties of each CV subsample, then discuss the CRTS completeness in Section~\ref{sec:alllegacy}.

\subsection{The SDSS Legacy Survey} \label{sec:legacy}
The Sloan Digital Sky Survey \citep[SDSS;][]{sdsstech_york00} is a large multicolour imaging and spectroscopic survey, carried out using a dedicated 2.5\,m telescope at Apache Point Observatory (APO) in New Mexico, USA. The imaging survey was completed and released in 2011 as Data Release 8 \citep[DR8;][]{sdssdr8}. It included $ugriz$ photometry of more than 900 million objects, covering 14\,555 square degrees of sky.

A set of complex criteria govern the spectroscopic target selection process. The spectroscopic survey is carried out as a number of specific, targeted programmes, each with its own photometric colour selection algorithms and exclusion boxes in colour space. Currently in the third phase of the survey (SDSS-III), the focus is on measuring the cosmic distance scale with the Baryonic Oscillations Spectroscopic Survey (BOSS), but previous phases included e.g. the Sloan Supernova Survey, the Sloan Extension for Galactic Understanding and Exploration (SEGUE) and the Multi-object APO Radial Velocity Exoplanet Large-area Survey (MARVELS). 

The main survey from SDSS-I is known as the {\em Legacy Survey}. It employed a well-defined, uniform targeting strategy with the aim to produce complete, magnitude-limited samples of galaxies, luminous red galaxies and quasars \citep[][respectively]{sdssgal_strauss02,sdsslrg_eisenstein01, sdssqso_richards02}. 
The survey covered 7646 square degrees of the North Galactic Cap, and later also three stripes (740 square degrees in total) in the South Galactic Cap. Data release 7 (DR7) provided the complete data release for the Legacy survey \citep{abazajian09_sdssdr7}.

The equatorial stripe in the South Galactic Cap is known as Stripe 82. It was observed repeatedly throughout SDSS-I and SDSS-II, in order to identify variable objects and, through co-addition of the images, provide photometry up to 2 magnitudes deeper than the North Galactic Cap region. Similarly, the spectroscopy taken in this part of the sky did not follow the standard selection criteria, but instead were used for various tests, such as new selection criteria, or completeness estimates of the small, deep samples \citep[e.g.][]{sdss_vandenberk05}. The data from Stripe 82 therefore does not have the same statistical properties as the rest of the Legacy sample, so great care should be taken when constructing statistically representative samples from the Legacy data. For this reason, in the rest of this paper, we will refer to the Legacy footprint as only the original contiguous survey ellipse of the Northern Galactic Cap as defined in \citet{sdsstech_york00}. The region is indicated as a grey shaded area in Figure~\ref{fig:footprint}. Of the 1043 CVs identified by CRTS, 221 (21 per cent) fall within the SDSS Legacy footprint.

\subsection{The SDSS CV sample} \label{sec:sdsscvs}

Cataclysmic Variables overlap with quasars in colour space, so some of the targets which were selected for spectroscopy in the Quasar Survey turned out to be CVs. \citet{szkody02sdssI,szkody03sdssII,szkody04sdssIII, szkody05sdssVI,szkody06sdssV,szkody07sdssVI,szkody09sdssVII,szkody11} produced a catalogue of 285 CVs identified in this way. The Quasar Survey is estimated to be 95 per cent complete down to $i<19.1$  \citep{sdss_vandenberk05,sdss_richards06}. The same selection criteria applies to the CV spectra, so it is expected that the completeness of the SDSS CV sample is similar. It is considered the most homogeneously selected, deepest sample of CVs, and as a result, it was the first in which the long-predicted period minimum spike was seen. \citet{gaensicke09} also showed that the composition of the CV population changes towards fainter apparent magnitudes. An increasing fraction of faint CVs have white dwarf dominated optical spectra, suggesting that the majority of the faint CVs are intrinsically faint (due to their low accretion rates), rather than simply because they are more distant. Since CRTS can detect CVs that are several magnitudes fainter than the spectroscopic limit of SDSS it is of particular interest to compare the CRTS CVs to those discovered by SDSS.

\begin{figure*}
\centering
\rotatebox{0}{\includegraphics[width=15.0cm]{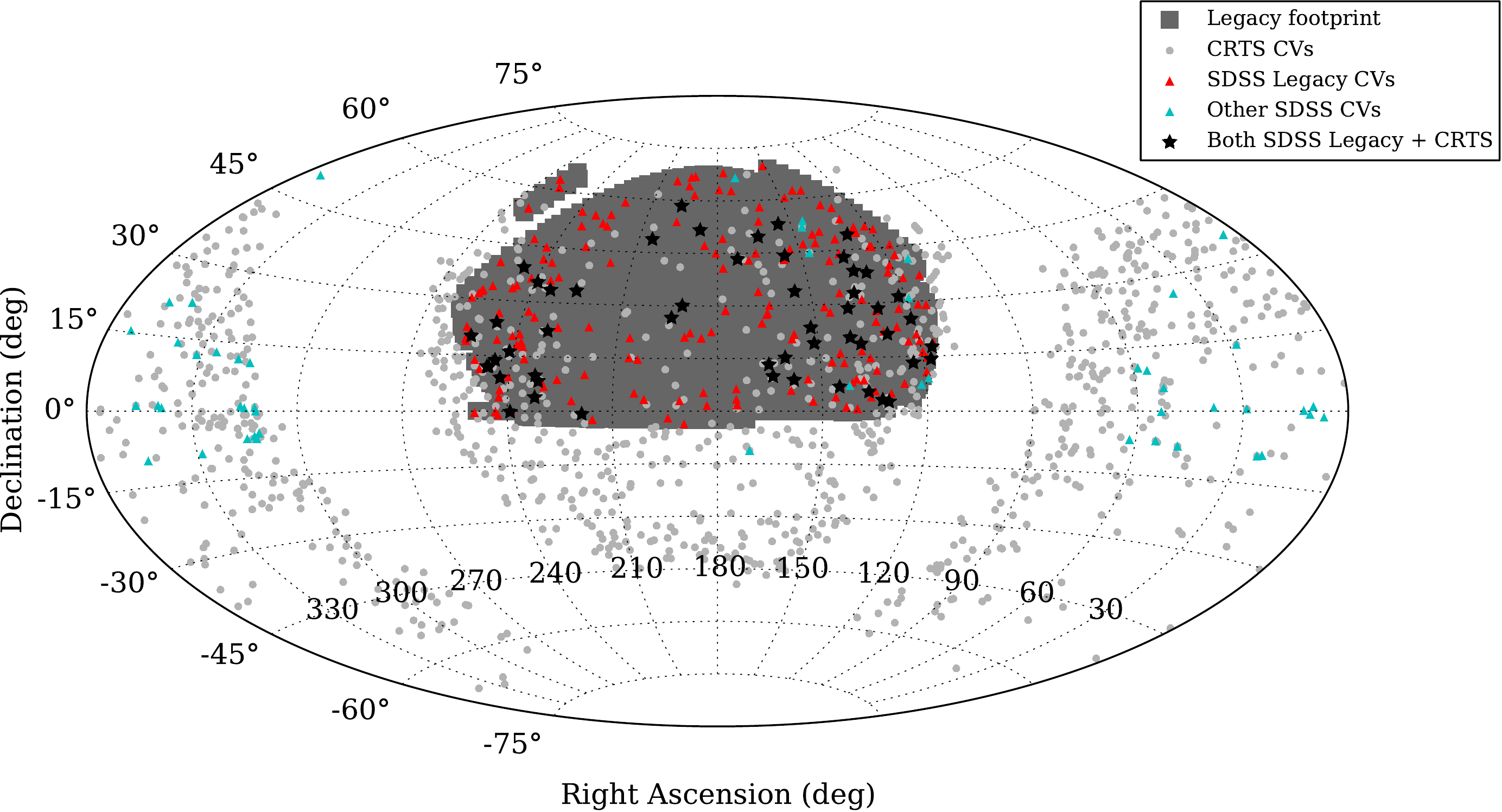}}
~\\~\\
\rotatebox{0}{\includegraphics[width=15.0cm]{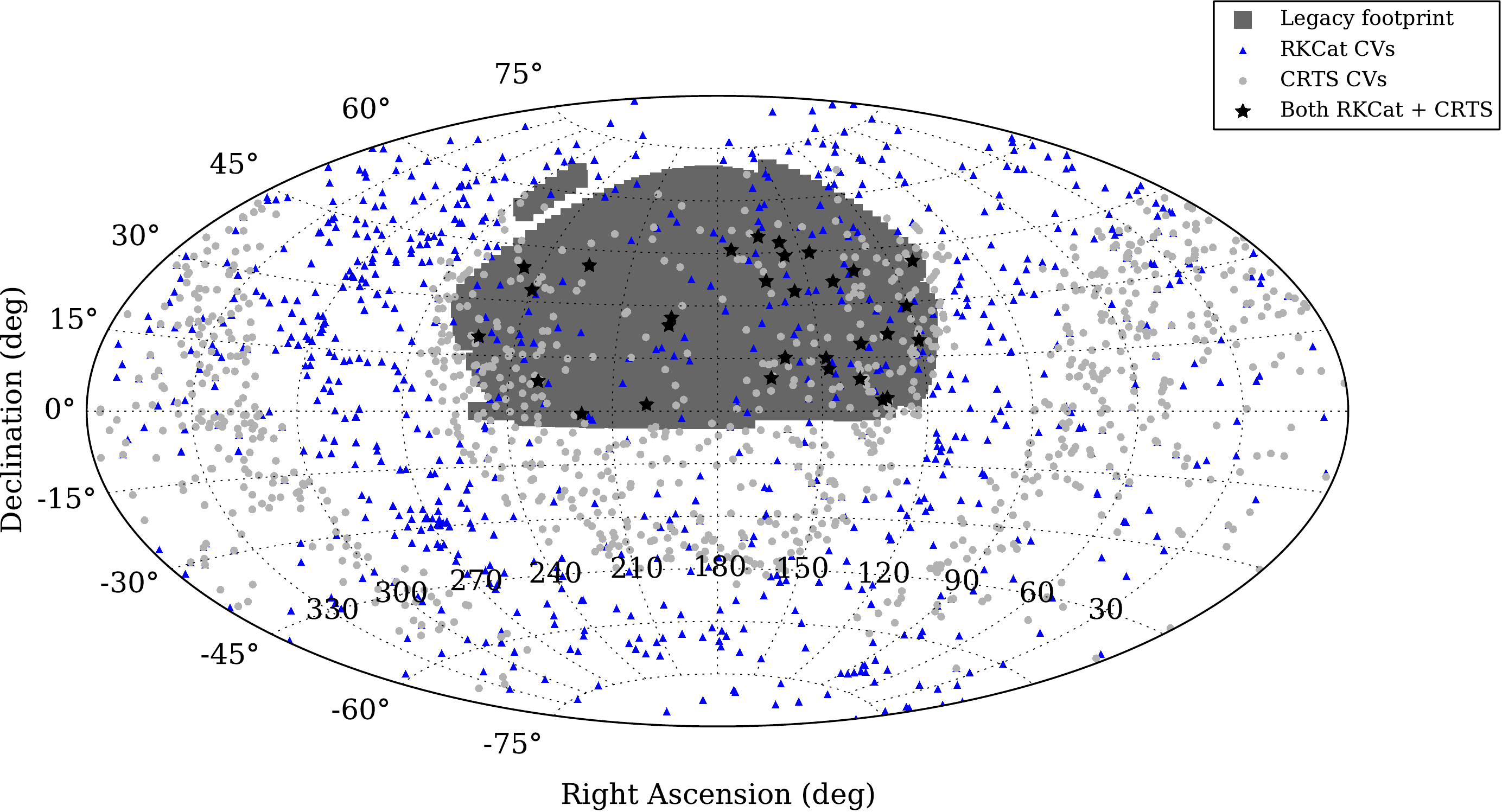}}
\caption{\label{fig:footprint} An all-sky comparison of the CRTS and SDSS samples of CVs (top) and the CRTS and Ritter-Kolb CVs (bottom), plotted in equatorial coordinates. The 1043 CVs detected by the three telescopes of CRTS are shown as light grey dots and the SDSS Legacy footprint is shaded in dark grey. A total of 221 CRTS CVs are located within the Legacy footprint. The red triangles in the top plot show the locations of the 219 spectroscopically selected SDSS Legacy CVs and the black star symbols in that plot represent the 52 SDSS Legacy CVs which were re-identified by CRTS. For completeness, we also show the 50 SDSS CVs which fall outside of the Legacy footprint, or were identified by surveys other than the Legacy Quasar Survey. They are shown as light blue triangles. Similarly, the bottom plot shows the the 862 CVs from the Ritter-Kolb catalogue in dark blue triangles along with the CRTS CVs and the Legacy footprint. Again, the black star symbols indicate the 30 CVs in the Legacy footprint which were discovered by surveys or observations other than CRTS, but subsequently re-identified by CRTS.} 
\end{figure*}

As was already pointed out by \citet{thorstensenskinner12}, not all 285 targets in the \citet{szkody11} catalogue are in fact CVs. \citet{thorstensenskinner12} label 14 systems as non-CVs, and we additionally exclude SDSSJ061542.52+642647.6 and SDSSJ202520.13+762222.3 which are poorly studied systems and unconfirmed as CVs. Their spectra resemble K-type stars with weak \halpha\, emission, but no other Balmer emission lines. Both stars were targeted as part of SEGUE, and their photometric colours make them more likely to be giant stars with chromospheric activity rather than CVs. This leaves 269 CVs in the sample, 228 of which fall within the Legacy footprint. However nine of these spectra were not obtained as part of the Legacy program, but were selected using the target algorithms for SEGUE and MARVELS. We exclude these as well, leaving 219 SDSS Legacy CVs in the sample. The SDSS CV sample is shown along with the CRTS CVs in Figure~\ref{fig:footprint}, relative to the sky footprint of the Legacy survey. 

Only 52 of the 219 SDSS Legacy CVs (23.7 per cent) were re-identified by CRTS as a result of their optical variability. We downloaded the CSDR2 light curves (see Section~\ref{sec:crtslcs}) of all the SDSS Legacy CVs and visually inspected them to understand the reasons they were not detected as CVs by CRTS. The results are summarised in Table~\ref{tab:recovery}. 
As expected, most of the SDSS CVs missing from CRTS are systems which show very little photometric variability, or variability with an amplitude less than 2~magnitudes. Some of the CVs which show `little variability' will be high accretion rate `novalike' systems, which have hot, stable discs which never undergo outbursts. Others, in this same category, are likely to be dwarf novae, but no outbursts have so far been observed by CRTS. Future CRTS observations may still recover some of these dwarf novae. The 2~magnitude (or 3$\sigma$) threshold used to select transients (see below) excludes CVs which display photometric variability smaller this limit, but this cut is necessary to avoid including too many large amplitude variable stars in the sample of transients. This cut seems to affect the polars in particular, since the state transitions increase the observed scatter in the light curve and are also typically $\Delta m\lesssim2$ mag. 

CRTS covers almost all of the Legacy footprint, but there are some small gaps in RA at northern declinations, $\delta\gtrsim55$\degr. Two SDSS CVs (0.9 per cent) fall outside the CRTS coverage area for this reason. Only three of the SDSS CVs (1.4 per cent) have sparsely observed light curves, which suggests that coverage is not the main cause of missing known CVs. 

The pipeline photometry from each night is compared both to past CRTS observations (using a minimum of 40 images, median stacked), as well as to photometric catalogues such as the USNO\footnote{United States Naval Observatory. Photometric catalogue  available from http://tdc-www.harvard.edu/catalogs/ub1.html (USNO-B1.0)} or SDSS catalogues. 
In order to be flagged as a transient, a target has to be more than 2~mag brighter than the catalogue magnitude, or brighter than three times the scatter in the historical light curve of the target, with a minimum brightening of 0.65~mag \citep{drake14}. It is not possible to flag targets which are measured to be much fainter than their comparison magnitudes, because poor weather conditions and chip defects would cause too many spurious detections. Therefore, if a CV is in outburst or in a high state in the comparison catalogue, it is unlikely to ever meet the criteria to be flagged as a transient. This accounts for a significant fraction of the SDSS CVs which were not re-identified by CRTS. CRTS maintains a CV {\em Watchlist}\footnote {http://nesssi.cacr.caltech.edu/catalina/CVservice/CVtable.html}, where known CVs which are in outburst in comparison catalogues are monitored. 33 SDSS CVs (15.1 per cent) appear on the {\em Watchlist}. It is unclear how many CVs may be undiscovered as a result of an outburst magnitude in the comparison catalogue, but it exposes a bias against systems with very high accretion rates and frequent outbursts, and polars with extended high/low states, as they are more likely to be bright in the comparison catalogues. 

A further three SDSS CVs (1.4 per cent) have been detected, but their classification as CVs were not certain enough to be added to the CV list. Transients like these, which are detected in a high/outburst state, but are not detected in the median stacked image, are added to the {\em Orphancat} database and may need additional observations to secure their classifications. All three of the SDSS CVs which appear only in {\em Orphancat} have $g>20$ in quiescence.

In summary, 23.7 per cent of SDSS Legacy CVs are recovered by CRTS. Those that were not identified can be explained within the CRTS photometric limits and selection criteria. 


\begin{table*}
\caption{\label{tab:recovery} CRTS recovery of CVs from the SDSS and Ritter-Kolb catalogues.}
\begin{tabular}{lrrrrr}
\hline
 ~ & \multicolumn{2}{c}{SDSS} & ~ & \multicolumn{2}{c}{RKCat} \\
\hline\hline
\smallskip
  Total number of Legacy CVs:                    & \multicolumn{2}{c}{219} &~& \multicolumn{2}{c}{129}  \\ 
\smallskip
  Recovered by CRTS:                             & 52 & (23.7\%) &~&   30 & (23.3\%) \\ 
  Not recovered:                                 &  ~ &  ~       &~&    ~ & ~ \\ 
  \phantom{X}Little photometric variability      & 80 & (36.5\%) &~&   29 & (22.5\%)\\ 
  \phantom{X}Variable, but not transient, or var$<2$mag & 42 & (19.2\%) &~& 24 & (18.4\%) \\
  \phantom{X}Bright in comparison catalogue      & 33 & (15.1\%) &~&   26 & (20.2\%)\\ 
  \phantom{X}Saturates in outburst               &  0 & (0\%)    &~&    9 & (7.0\%)\\ 
  \phantom{X}Unreliable photometry/blending      &  4 & (1.8\%)  &~&    6 & (4.7\%)\\ 
  \phantom{X}Orphancat: too faint in quiescence  &  3 & (1.4\%)  &~&    1 & (0.8\%)\\ 
  \phantom{X}Poor coverage                       &  3 & (1.4\%)  &~&    3 & (2.3\%)\\ 
  \phantom{X}Area not covered by CRTS            &  2 & (0.9\%)  &~&    0 & (0\%)\\       
  \phantom{X}Should be detected -- outbursts     &  0 & (0\%)    &~&    1 & (0.8\%)\\ 
\hline                                                          
\end{tabular}                                                  
\end{table*}

\subsection{The Ritter-Kolb catalogue} \label{sec:rk}

We carried out a similar analysis on the Ritter-Kolb catalogue of CVs \citep[RKCat,][]{rkcat}\footnote{http://www.mpa-garching.mpg.de/RKcat/}. 
This catalogue contains all CVs that are studied well enough that at least their orbital periods are known, and lists various properties of each CV. 
It is the only sample that covers low galactic latitudes and the galactic plane. The latest edition, version 7.20, contains 1094 entries, including several CVs which were discovered by CRTS and SDSS. We first filter out these systems from the catalogue so that it provides an independent sample of 862 CVs, and then select those which lie within the Legacy footprint to compare to the CRTS and SDSS Legacy samples. For simplicity, we will use the term `Ritter-Kolb CVs' in this paper to refer only to those CVs which were discovered by methods or surveys other than CRTS or SDSS.  

Figure~\ref{fig:footprint} shows the sky distribution of the 862 Ritter-Kolb CVs along with the full CRTS sample. 129 Ritter-Kolb CVs fall within the Legacy footprint, of which 30 systems (23.3 per cent), known prior to the start to CRTS, were recovered. 

As for the SDSS sample, the greatest reason for not recovering more Ritter-Kolb CVs, is a lack of large amplitude photometric variability (Table~\ref{tab:recovery}). 55 targets in the sample (41 per cent) display variability of less than 2~mag, or variability that does not meet the transient selection criteria. Outburst magnitudes in comparison catalogues account for 26 of the missing systems (20.2 per cent), all of which are included in the {\em Watchlist}. Additionally, the Ritter-Kolb CVs suffer from saturation problems in CRTS. They are typically brighter in quiescence than the SDSS and CRTS CVs (see Figure~\ref{fig:maghisto}) so when observed in outburst, they will saturate ($V\sim12$) and will be rejected by the transient pipeline. In fact, CSDR2 cautions about the reliability of photometry $V<13$, because of the nonlinearity of the detector response at these magnitudes. We found nine such CVs in the sample (7.0 per cent). 
Six CVs (4.7 per cent) are rejected because of blending or unreliable photometry. Two of these reside in clusters (M5-V101 in the globular cluster M5 -- \citealt{hourihane11}, and the polar EU\,Cnc inside the open cluster M67 -- \citealt{williams13}) making it difficult to reliably identify the transient object in the low resolution CRTS images. CRTS now excludes a small region around every known globular cluster to reduce spurious detections as a result of crowding.
Three CVs have a nearby bright ($V<14$) star which affects the photometry, and one other is blended with a fainter nearby star. 

In total, 23.3 per cent of the Ritter-Kolb Legacy sample is recovered by CRTS.
We found only one CV which should have been detected but was missed for unknown reasons. This CV, SDSS\,J075107.50+300628.4, was previously identified as a dwarf nova by \citet{wils10}, and its CRTS light curve displays multiple bright outbursts. It is not a {\em Watchlist} CV.

\subsection{Completeness of the CRTS CV sample} \label{sec:alllegacy}

\begin{figure}
\centering
\rotatebox{270}{\includegraphics[width=5.2cm]{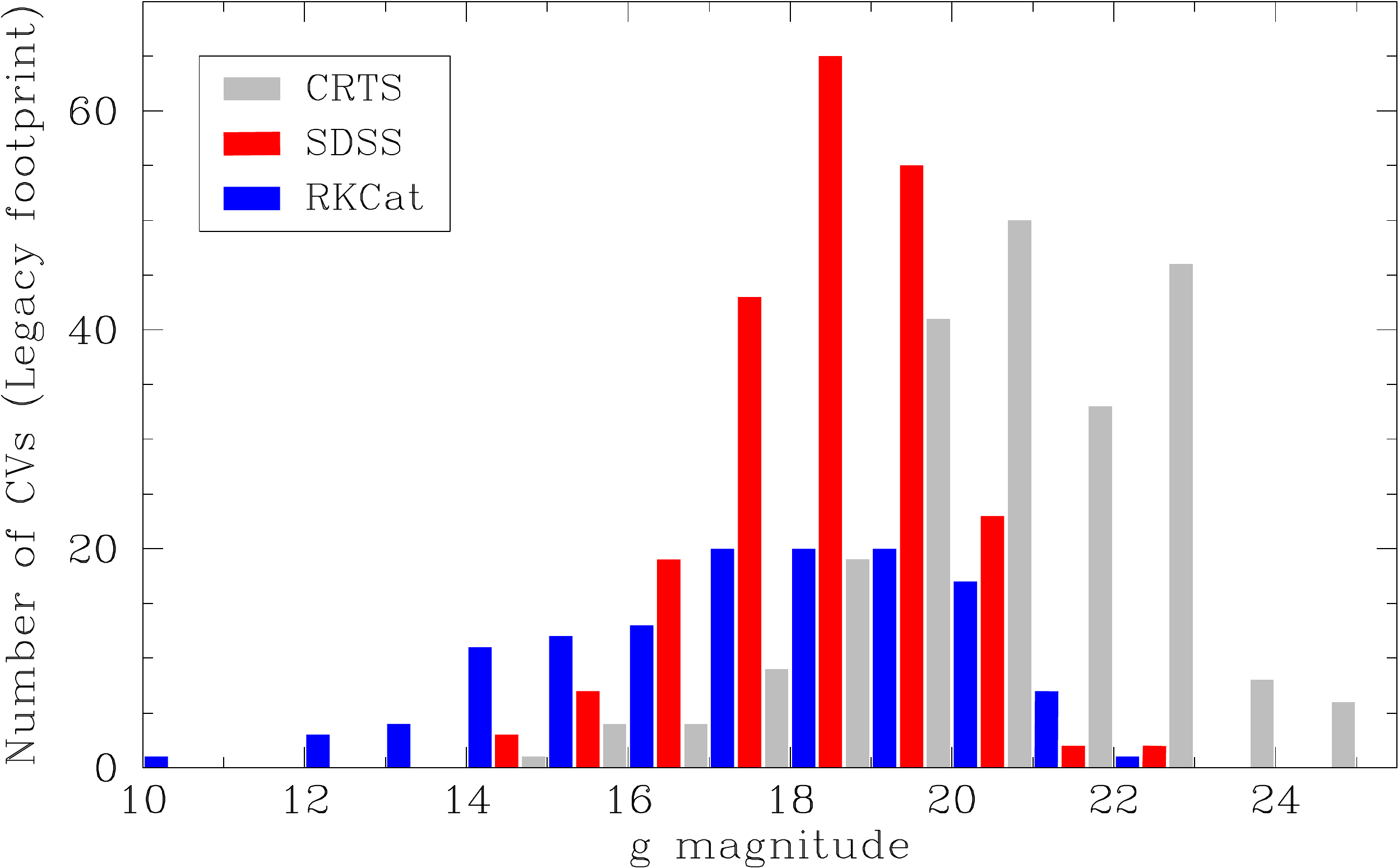}}
\caption{\label{fig:maghisto} SDSS-$g$ band magnitude distribution of the SDSS (red), Ritter-Kolb (blue) and CRTS (grey) samples of CVs in the Legacy footprint. CRTS contributes most of the CVs fainter than $g=20$ to the known population.} 
\end{figure}

As discussed in the previous two sections, CRTS recovered 23.6 per cent of the known SDSS+RKCat CVs in the Legacy footprint. 
Although this may seem low, it is important to remember that since CRTS is sensitive only to CVs which display large amplitude variability, its overall {\em external} completeness is not expected to be high. Table~\ref{tab:recovery} shows that between $\sim40-56$ per cent of CVs are hidden from transient surveys due to their lack of large amplitude ($>2$mag) photometric variability.
The {\em internal} completeness of the survey appears to be high. We found only one of the $(219+129=)348$ known CVs in this part of the sky which was not detected by CRTS and which could not be explained by the transient selection criteria and photometric limits of the survey (Table~\ref{tab:recovery}).  

If we restrict the analysis only to dwarf novae in the Legacy samples, we find that 144/219 of the SDSS CVs are classified as dwarf novae, of which 47 (32.6 per cent) were recovered. The dwarf nova recovery fraction is similar in RKCat. 70/129 CVs are classified as dwarf novae, and 26 (37.1 per cent) of these were recovered.  
\citet{thorstensenskinner12} suggest that a bias towards large outburst amplitudes could account for the known CVs which were not detected. However, we find that the light curves of most of the `missing' dwarf novae fall into the `Little variability' category, i.e. even though previous observations have shown these systems to be dwarf novae, no outbursts have been observed by CRTS over the six years of the project. The outburst frequency and observing cadence therefore also play a major role in the detection probability of a given dwarf nova. 23 of the `missing' dwarf novae are included on the {\em Watchlist}. 

\begin{figure}
\centering
\rotatebox{270}{\includegraphics[width=5.3cm]{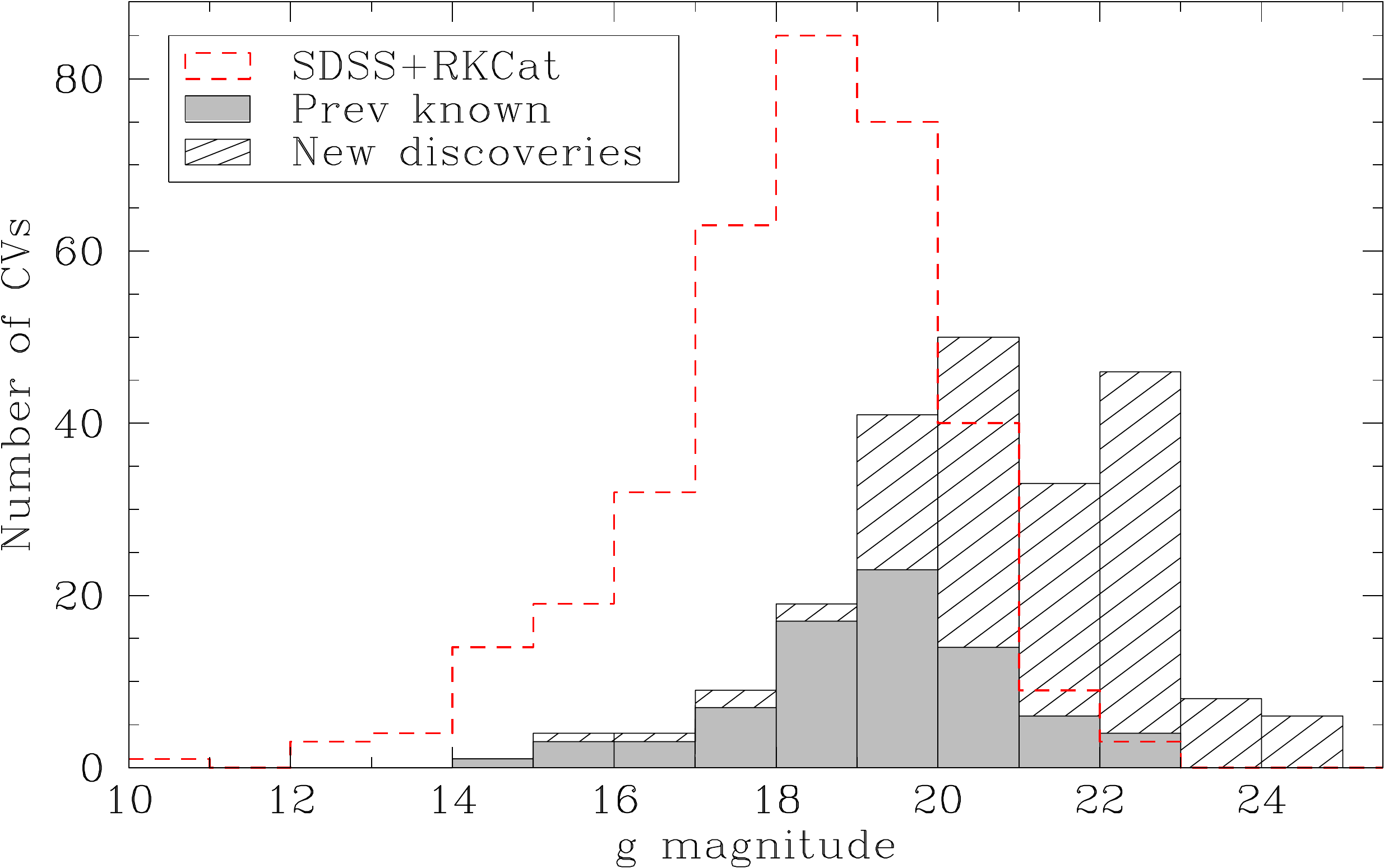}}
\caption{\label{fig:oldnew} $g$ band magnitude distribution of the recovered, previously known CVs (grey) and newly discovered CVs (black hatched) in CRTS. The sample of previously known CVs (SDSS+RKCat) is shown for comparison (red dash). Note the large number of new CVs in the range $19<g<23$.} 
\end{figure}

Both \citet{wils10} and \citet{thorstensenskinner12} have already noted that most of the new CVs discovered by CRTS are faint systems that could not be detected in quiescence by SDSS or other previous surveys. This is illustrated again in Figure~\ref{fig:maghisto}, which shows a direct comparison between the quiescent $g$ band magnitude distributions of the three Legacy samples of CVs. CVs collected in the Ritter-Kolb catalogue dominate the bright end of the distribution, where SDSS and CRTS saturate. On the faint side, CRTS is responsible for most of the CVs with $g\geq20$ and almost all fainter than $g\geq22$. With very little to compare the sample to, assessing the external completeness of CRTS at these faint magnitudes directly from observations is not possible. 
Instead, we investigate the sensitivity of the survey to faint CVs, by dividing the sample into `new discoveries' and `previously known' CVs (Figure~\ref{fig:oldnew}). The $g$ band distribution of the combined SDSS+RKCat Legacy sample is shown as a red, dashed-line histogram, representing all CVs known before the start of CRTS. The CVs recovered by CRTS are shown as a grey filled histogram. 
Very few CVs brighter than $g<18$ were recovered --- CRTS is much more efficient at detecting the fainter systems. The number of newly discovered CVs (indicated by the hatched bars) rapidly increases for $g>19$ and drops sharply at $g>23$. Within this range CRTS recovered 47 known CVs, but it also discovered 123 new CVs, 2.6 times the number of known CVs it recovered. This suggests that the survey is particularly efficient at finding CVs with quiescent magnitudes between $19\lesssim g\lesssim23$. This is probably a consequence of the dynamic range of the detectors and the depth of the comparison surveys: a definite detection can be made in quiescence, but a bright outburst can also be accommodated without saturating the detector. 

The sharp drop in the number of systems with $g>23$ is however a feature of the CV population rather than a limit to the survey sensitivity. SDSS photometry is available for the whole of the Legacy footprint, complete to a depth of 22.2 in the $g$ band \citep{sdssdr2}. Slightly deeper photometry is available in areas observed under good seeing and low sky brightness conditions. If a substantial part of the CV population had apparent magnitudes fainter than $g\geq23$, we would expect to find many transients with no counterpart in the SDSS photometric catalogue. However, of all the transients in the Legacy footprint, we found only one which does not have a corresponding SDSS object. This must mean that $g\sim23$ represents a distance limit intrinsic to the CV population, rather than simply the limiting magnitude of the SDSS photometry or the photometric limit set by the fixed exposure time of the surveys used by CRTS. Assuming $M_g=11.6$ as the average absolute magnitude of low-luminosity CVs \citep{gaensicke09}, $g=23$ corresponds to a distance of $d=1905$~pc. Even considering the range of Galactic latitudes covered by the Legacy footprint, this suggests that the CV population is distributed well beyond the Galactic thin disc. A more detailed analysis of the scale height of the CRTS CV population is presented by \citet{drake14}, who also conclude that many of the CRTS CVs belong to the Galactic thick disc.

\subsection{Spectroscopic completeness and the white dwarf dominated CVs} \label{sec:wddom}

\begin{figure}
\centering
\rotatebox{270}{\includegraphics[width=8.0cm]{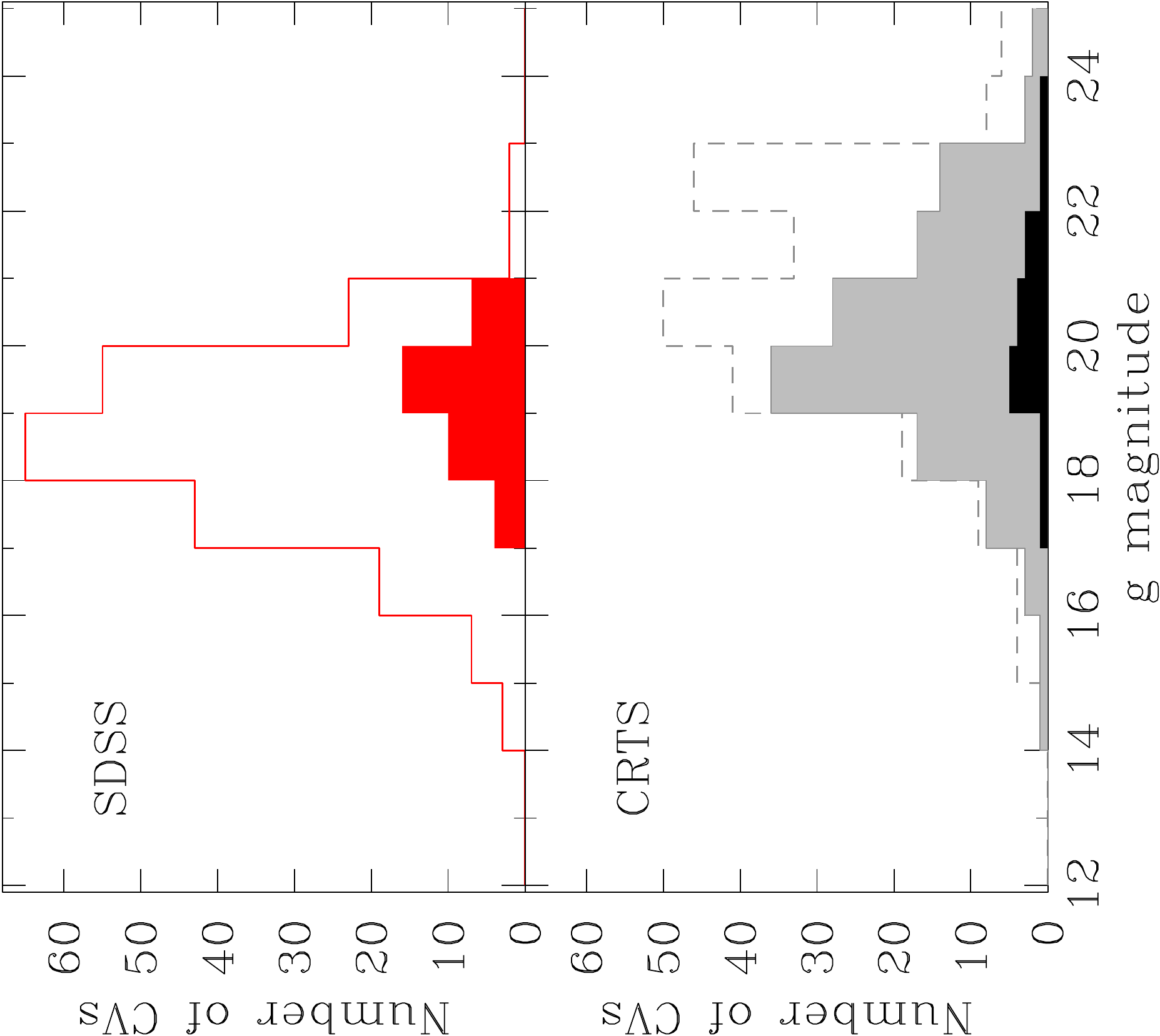}}
\caption{\label{fig:wdhisto} $g$ band magnitude distribution of the SDSS (top) and CRTS (bottom) CVs in the Legacy footprint. Spectra are available for all the SDSS CVs. Those which are white dwarf dominated are indicated by the filled red bars. The full CRTS Legacy sample is shown as a dashed grey histogram, and those for which spectra are available as filled grey bars. The white dwarf dominated CRTS CVs are shown in black.} 
\end{figure}

One of the key features of the SDSS CV sample was a large number of systems with white dwarf dominated spectra and short orbital periods, $\Porb\lesssim100$~min \citep{gaensicke09}. All of these CVs are fainter than $g=17.4$ and display no features from the secondary star in their spectra, so these are likely to be evolved, low accretion rate, WZ\,Sge-type dwarf novae.
We now look for similar CVs in the CRTS sample. The bottom panel of Figure~\ref{fig:wdhisto} illustrates the spectroscopic follow-up of the CRTS CVs within the Legacy footprint. The spectroscopic identification is essentially complete up to $g\leq19$, largely thanks to the SDSS spectroscopic survey. Other spectroscopic follow-up observations, including those presented in Section~\ref{sec:data}, have so far provided spectra for 45 per cent of CRTS Legacy CVs fainter than $g\geq20$. The CVs with spectra showing the wide absorption wings characteristic of a white dwarf are shown as black solid bars. As in the case of the SDSS, these WZ\,Sge candidates are a small fraction of the total number of spectra available ($33/219=15.1$~per~cent for SDSS, $16/130=12.3$~per~cent for CRTS). 

Four of the 33 SDSS white dwarf dominated CVs shown in Figure~\ref{fig:wdhisto} were recovered by CRTS (12.1 per cent). Such a small recovery fraction is consistent with the interpretation that they are low accretion rate systems with long outburst recurrence times. The CRTS light curves of the four systems which were recovered also only display a small number of outbursts. They are SDSS\,J133941.12+484727.4 (= V355\,UMa, 1 outburst), SDSS\,J161027.61+090738.4 (1 outburst), SDSS\,J155644.23-000950.2 (= V493\,Ser, 2 outbursts) and SDSS\,J122740.82+513924.9 (3 outbursts). The four CVs fall one each in the four magnitude bins from $g=17-21$. The 12 new contributions from CRTS display similarly few outbursts: two with three outbursts, three with two outbursts and seven with a single outburst. 

Three outbursts in six years, as observed in some of these white dwarf dominated CVs, imply a much higher accretion rate than in WZ\,Sge itself, for which the outburst recurrence time is 2--3 decades. CRTS is only starting to uncover the low accretion rate population, many more low accretion rate CVs remain to be discovered (Figure~\ref{fig:cumulative_outb}). 
In contrast to the SDSS sample, which showed that an increasing fraction of CVs towards fainter magnitudes are white dwarf dominated \citep{gaensicke09}, we find that the white dwarf dominated 
spectra represent a constant fraction ($\sim$14 per cent) of the spectra in each magnitude bin $g>19$. We recognise however that our spectroscopic follow-up is subject to a bias, since proportionally fewer low accretion rate objects were discovered early in the survey when our spectroscopic observations were carried out. 
The 6 year length of the CRTS project is still short compared to  the long outburst recurrence times of WZ\,Sge dwarf novae. For example, 25/33 (75.8 per cent) of the SDSS Legacy CVs which are white dwarf dominated show only small amplitude variability in their CRTS light curves and have no observed outbursts\footnote{33 of the SDSS Legacy CVs are white dwarf dominated. 4 are recovered by CRTS, 2 are {\em Watchlist} targets, 2 have poor CRTS coverage and 25 show only small amplitude variability.}.  

CV population models predict that $\sim50-77$ per cent of all CVs should have evolved beyond the period minimum \citep[e.g][]{kolb93,howell01,knigge11}, and WZ\,Sge-type dwarf novae are often considered the best candidates for this evolved population. However, even at the faint magnitudes probed by CRTS, observations cannot (yet) account for the large number of systems expected from theoretical models.

%
\section{Summary} \label{sec:summary}

We presented identification spectra of 85 transient objects identified as cataclysmic variable candidates by the Catalina Real-time Transient Survey. 
72 of the spectra are from our own follow-up program on the Gemini Telescopes and GranTeCan, seven are from the Sloan Digital Sky Survey Baryonic Oscillations Spectroscopic Survey (plus one additional spectrum of a target also observed by us on Gemini North) and a further six are from the Public ESO Spectroscopic Survey for Transient Objects. Our sample includes three AM\,CVn binaries, at least one helium-enriched CV, one polar and 80 dwarf novae. 

25 of the spectra display clear absorption features of the white dwarf in the Balmer lines, indicative of a low accretion rate. We found five CVs for which both the white dwarf and the donor star are visible in the spectrum, all with orbital periods at the lower edge of the period gap. The spectrum of CRTS\,J163805.4+083758 (=V544\,Her), shows evidence for a K-type star donor, but an (uncertain) period of only 100~min is reported for this system \citep{howell90}. Further observations are required to confirm this period, since a normal K star would be too large to fit in a 100~min binary. If the short period is confirmed, the donor star in this system must have been stripped of its outer layers during a rapid mass transfer phase earlier in its evolution \citep[see e.g.][]{thorstensen13}.

The spectra of 18 of our targets display double-peaked emission lines. Three of these are known eclipsing CVs and we also identified a new eclipsing system, CRTS\,J132536.0+210037, from its SDSS photometry. We were unable to determine the orbital period from its CRTS light curve, but the white dwarf features in its spectrum suggests that the period is $<90$~minutes. 
 
Using the CRTS light curves, we analysed the outburst properties of the full sample of 1043 CVs identified by CRTS during its first six years of operation. We showed that the rate at which new CVs are discovered by CRTS is slowing down. This is at least partially due to the fact that most of the high accretion rate systems were found early in the survey and that the lower accretion rate systems discovered now have a lower detection rate due to their longer outburst recurrence times. Low accretion rate CVs are expected to dominate future CRTS discoveries of dwarf novae, so the detection rate is expected to slowly decrease even further.

CVs with white dwarf dominated spectra have significantly fewer outbursts in their CRTS light curves than disc dominated systems. We do however find some white dwarf dominated CVs that have as many as 3 or 4 outburst during the six years of CRTS monitoring, which imply a much higher accretion rate than expected for WZ\,Sge systems. We found no significant difference between the outburst properties of disc dominated CVs and CVs at the lower edge of the period gap for which both the white dwarf and the donor star are visible in the spectrum.  

In order to estimate the completeness of the CRTS sample, we compared the CRTS CVs with known CVs from the SDSS CV and the Ritter-Kolb (RKCat) catalogues. In order to make a robust, unbiased comparison between the different samples, we selected only those CVs which fall within the SDSS Legacy footprint. This selection resulted in subsamples of 221/1043 CRTS CVs, 219/269 SDSS CVs and 129/862 RKCat CVs.
CRTS recovered 23.6 per cent of the SDSS+RKCat CVs. As expected, the main reason for CVs to be `missed' by CRTS (and by implication, other transient surveys as well) is the lack of large amplitude ($>2$mag) photometric variability. Dwarf novae with frequent outbursts or polars which undergo frequent transitions between high and low states are also more likely to be overlooked, because of the higher likelihood of being in outburst/high state in the comparison catalogues. The internal completeness
of CRTS is however high. We found only one out of the 348 known CVs in the Legacy footprint that was not detected and could not be explained by the transient selection criteria and/or photometric limits of the survey. 

CRTS is responsible for most of the known CVs fainter than $g>20$. There is a sharp drop in the number of systems for $g>23$, but we find only one CRTS CV in the Legacy footprint which does not have an SDSS photometric counterpart. This suggests that this limit is intrinsic to the CV population (i.e. a distance limit) and not simply a detection limit of the photometric survey. 

At present, spectra are available for 130/221 CRTS CVs in the Legacy footprint and they represent a constant fraction of the spectra in each magnitude bin $g>19$. Further spectroscopic identification of faint CRTS CVs should be carried out to determine whether the fraction of white dwarf dominated systems increase towards fainter magnitudes as observed in the SDSS CV sample \citep{gaensicke09}. The growing number of single-outburst CVs in the CRTS sample (Figure~\ref{fig:cumulative_outb}) suggests that CRTS is only starting to uncover the low accretion rate population, so many more CVs remain to be discovered, and future spectroscopic samples may be increasingly white dwarf dominated.


%
\section*{Acknowledgments}
We thank the referee, Steve Howell, for a helpful and positive report that improved the clarity of our presentation. 
We also thank Gayathri Eknath who helped to compile the outburst data used in Figure~\ref{fig:outbursthisto} during a summer project at the University of Warwick. 

EB and TRM are supported by a grant from the UK STFC, number ST/L000733/1. 
BTG is supported by the European Research Council under the European Union's Seventh Framework Programme (FP/2007-2013) / ERC Grant Agreement n. 320964 (WDTracer).
PRG is supported by a Ram\'on y Cajal fellowship (RYC--2010--05762), and acknowledges support provided by the Spanish grant AYA2012--38700. 
SGP and MRS acknowledge ﬁnancial support from FONDECYT in the form of grants number 3140585 and 1141269 respectively.
PS acknowledges support from the NSF through grant AST-1008734.

CRTS are supported by the U.S. National Science Foundation under grants AST-0909182 and CNS-0540369. The work at Caltech was supported in part by the NASA Fermi grant 08-FERMI08-0025, and by the Ajax Foundation. The CSS survey is funded by the National Aeronautics and Space Administration under Grant No. NNG05GF22G issued through the Science Mission Directorate Near-Earth Objects Observations Program.

The spectra presented in this paper were obtained with the Gemini Telescopes, under programs GN-2010B-Q-76, GS-2010B-Q-57, GN-2011A-Q-74, GS-2011A-Q-49 and GN-2013A-Q-89, and the Gran Telescopio Canarias (GTC), under program ID GTC30-11A.  
The Gemini Observatory is operated by the Association of Universities for Research in Astronomy, Inc., under a cooperative agreement with the NSF on behalf of the Gemini partnership: the National Science Foundation (United States), the National Research Council (Canada), CONICYT (Chile), the Australian Research Council (Australia), Minist\'erio da Ci\^encia, Tecnologia e Inova\c{c}\~{a}o (Brazil) and Ministerio de Ciencia, Tecnolog\'ia e Innovaci\'on Productiva (Argentina).
The Gran Telescopio Canarias (GTC) is operated at the Spanish Observatorio del Roque de los Muchachos of the Instituto de Astrof\'isica de Canarias, on the island of La Palma.

We gratefully acknowledge the use of the {\sc topcat} software \citep{topcat} in this research, the International Variable Star Index database (VSX) operated at AAVSO, Cambridge, Massachusetts, as well as the Sloan Digital Sky Survey (SDSS), funded by the Alfred P. Sloan Foundation, the Participating Institutions, the National Science Foundation, and the U.S. Department of Energy, Office of Science. Further information, including a list of participating institutions may be found on the SDSS web sites, http://www.sdss.org/ (for SDSS-I and SDSS-II) and http://www.sdss3.org/ (for SDSS-III).

\bibliographystyle{mn2e}
\bibliography{library4}

\bsp

%
\appendix

\section{Spectra and light curves of CRTS CVs}
In the figures that follow, we show the spectra of all 72 CRTS CVs we observed with Gemini and/or GTC, along with their CRTS lightcurves. Details of the observations are given in Table~\ref{tab:obslog} in the Online Materials.
The light grey vertical lines in the left hand panels indicate the rest wavelengths of the hydrogen Balmer lines (solid lines), the helium I and II lines (dashed lines) and the calcium triplet (dotted lines). The red dots are the quiescent $griz$ photometric measurements taken from SDSS DR7 and DR9. Where the SDSS photometry is not shown, the target was in outburst during our spectroscopic observations, or considerably fainter than the faintest SDSS photometry available. An asterisk following the target name indicates that the spectrum suffered blue flux loss due to differential atmospheric dispersion (see Section~\ref{sec:gemini}). The CRTS light curve of each target is shown in the right hand panel. For the nights when the target was observed but not detected, we plot upper limits (grey arrows), derived from nearby stars. The black horizontal dotted line is the $g$ band magnitude from SDSS. The figure captions give more information on specific targets where neccessary.

\setcounter{figure}{0}
\begin{figure*}
\centering
\rotatebox{270}{\includegraphics[height=17.5cm]{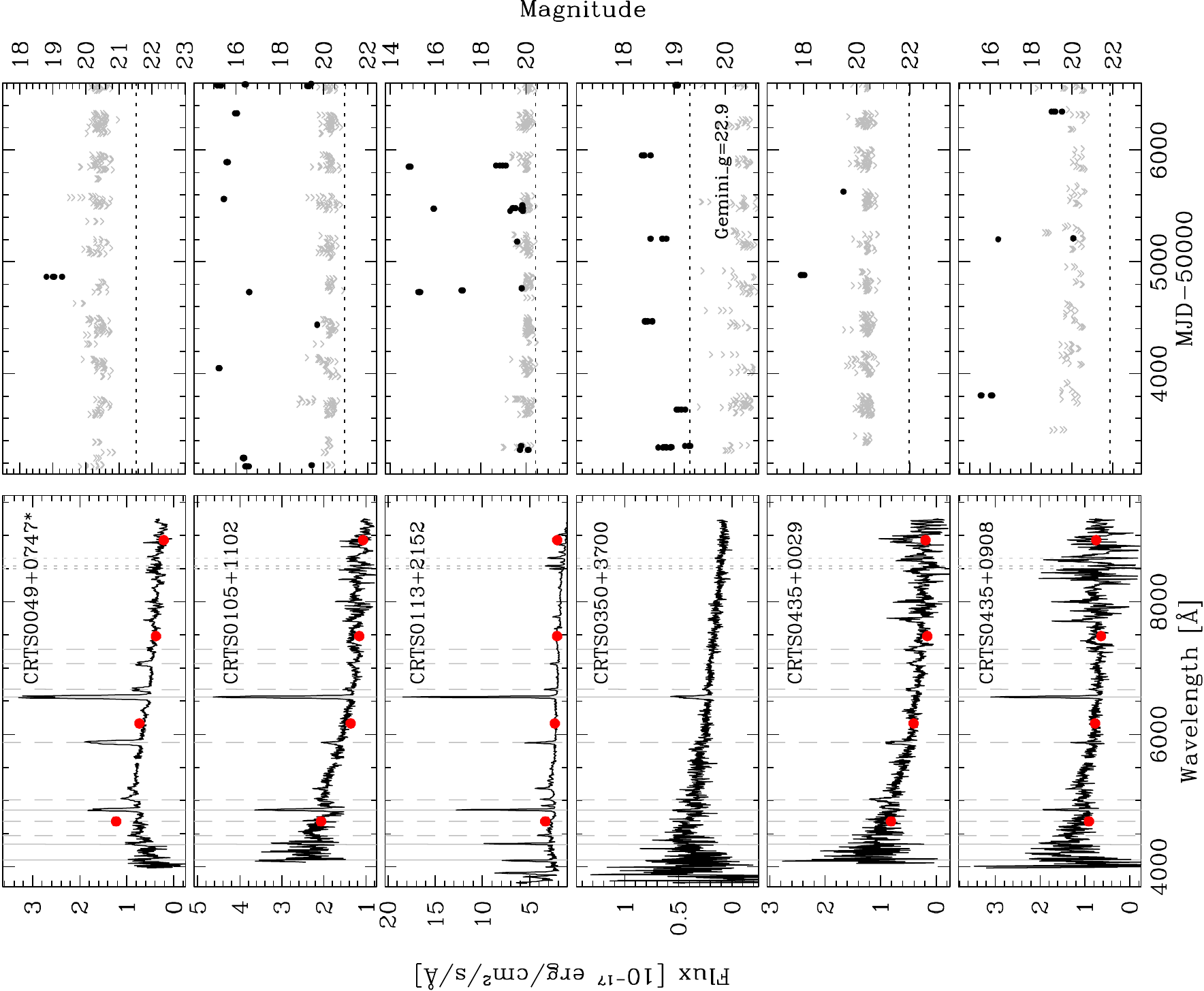}}
\caption{\label{fig:spectra} Spectra and light curves of CRTS CVs.  The SDSS photometry of CRTS\,J035003.4+370052 was taken during outburst. During our Gemini observation the target was much fainter, $g=22.9$. }
\end{figure*}
 
\setcounter{figure}{0}
\begin{figure*}
\centering
\rotatebox{270}{\includegraphics[height=17.5cm]{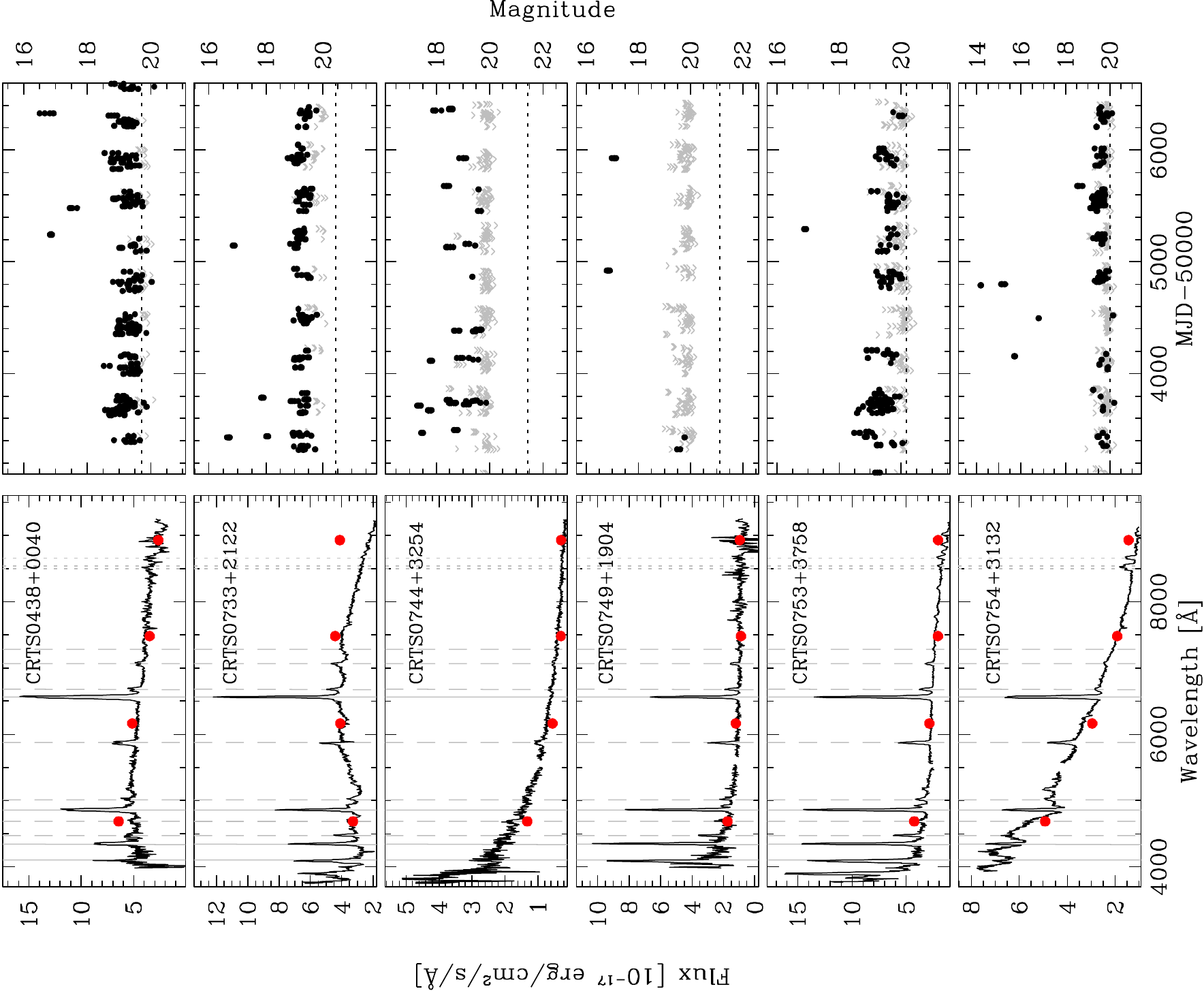}}
\caption{{\em continued.} Spectra and light curves of CRTS CVs.}
\end{figure*}
 
\setcounter{figure}{0}
\begin{figure*}
\centering
\rotatebox{270}{\includegraphics[height=17.5cm]{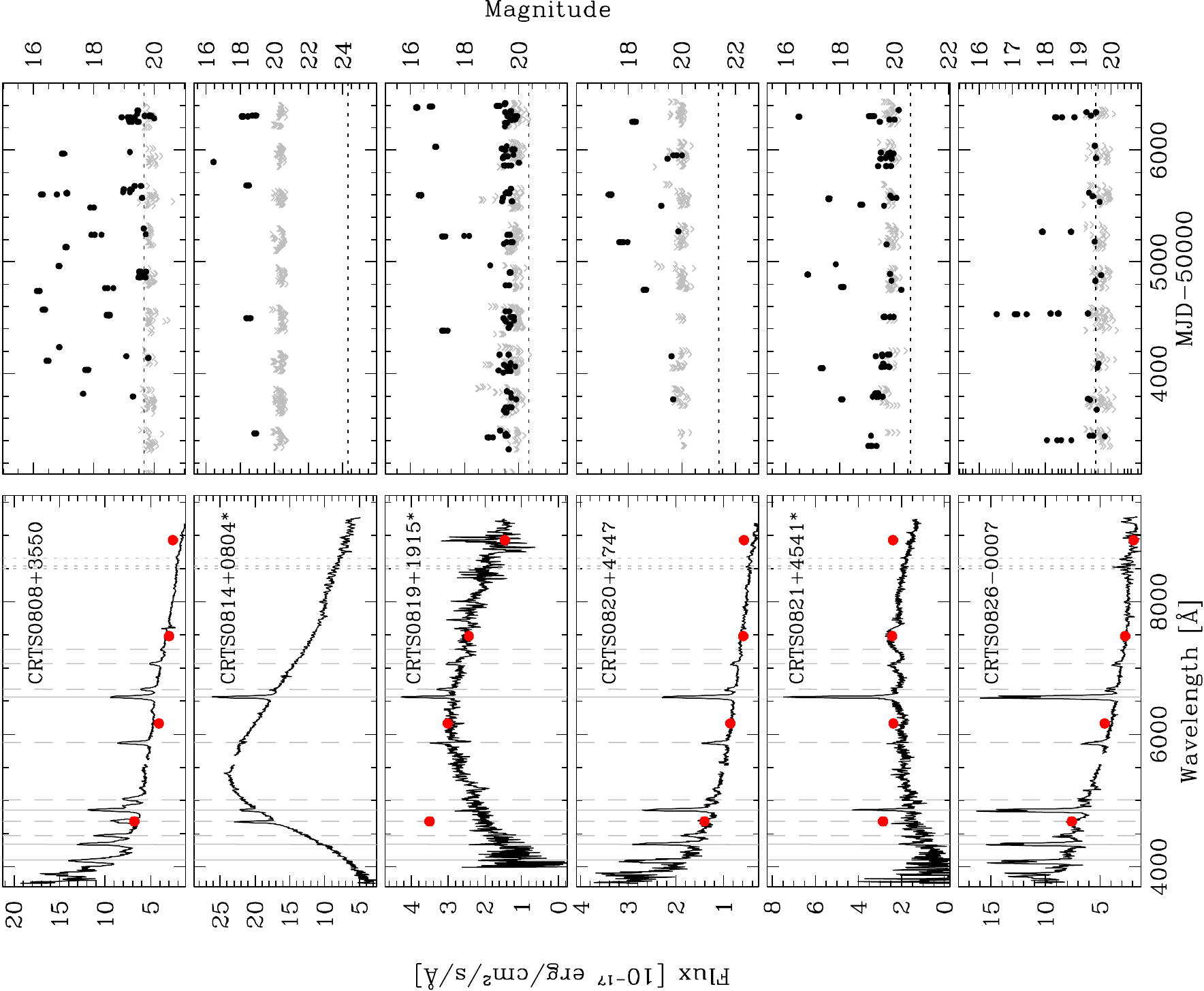}}
\caption{{\em continued.} Spectra and light curves of CRTS CVs. CRTS\,J081414.9+080450 was in outburst during our observations, so the SDSS photometry is not shown with the spectrum. The spectra of CRTS\,J081414.9+080450, CRTS\,J081936.1+191540 and CRTS\,J082123.7+454135 are affected by blue flux loss due to atmospheric dispersion (see Section~\ref{sec:gemini}).}
\end{figure*}
 
\setcounter{figure}{0}
\begin{figure*}
\centering
\rotatebox{270}{\includegraphics[height=17.5cm]{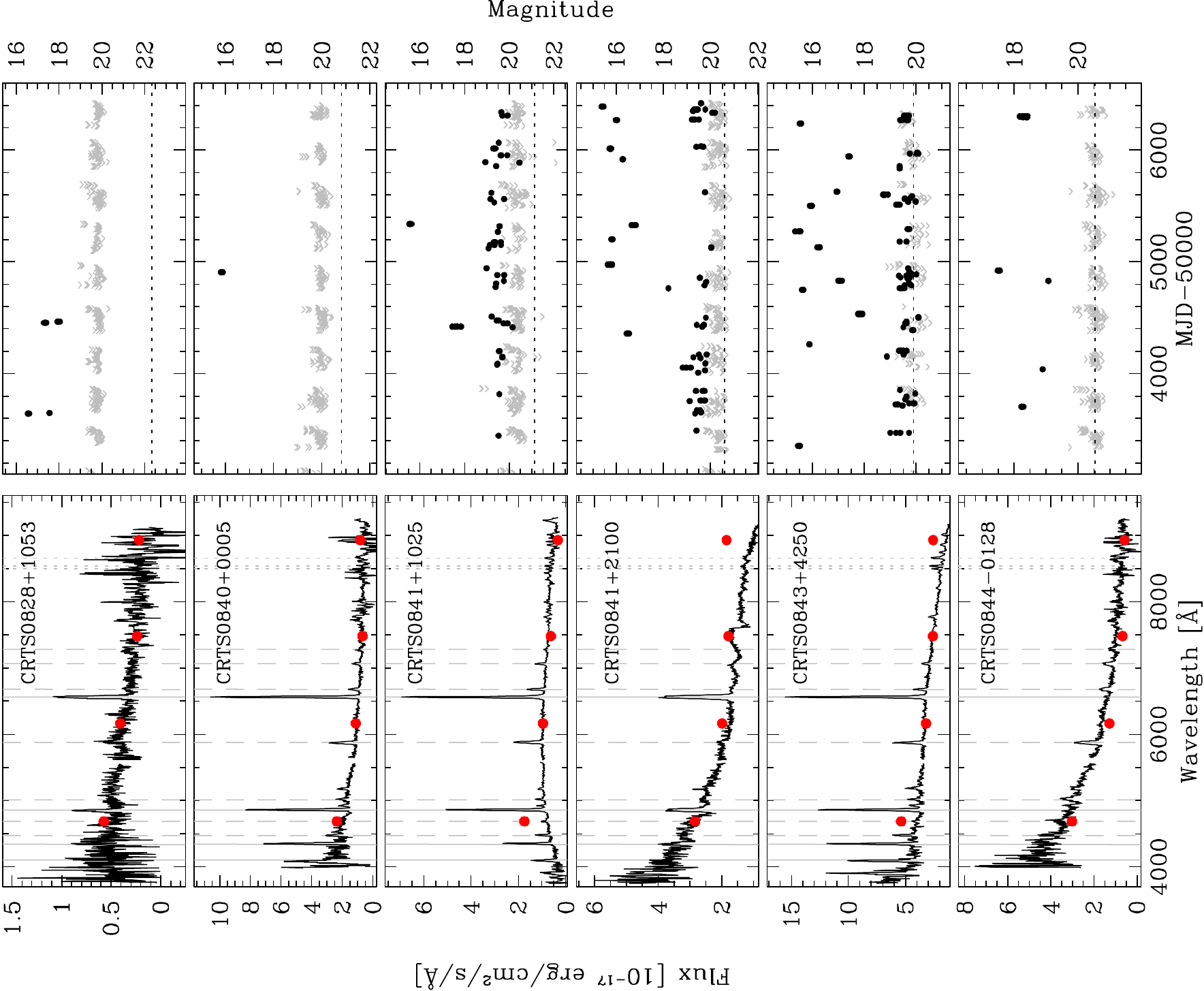}}
\caption{{\em continued.} Spectra and light curves of CRTS CVs.}
\end{figure*}
 
\setcounter{figure}{0}
\begin{figure*}
\centering
\rotatebox{270}{\includegraphics[height=17.5cm]{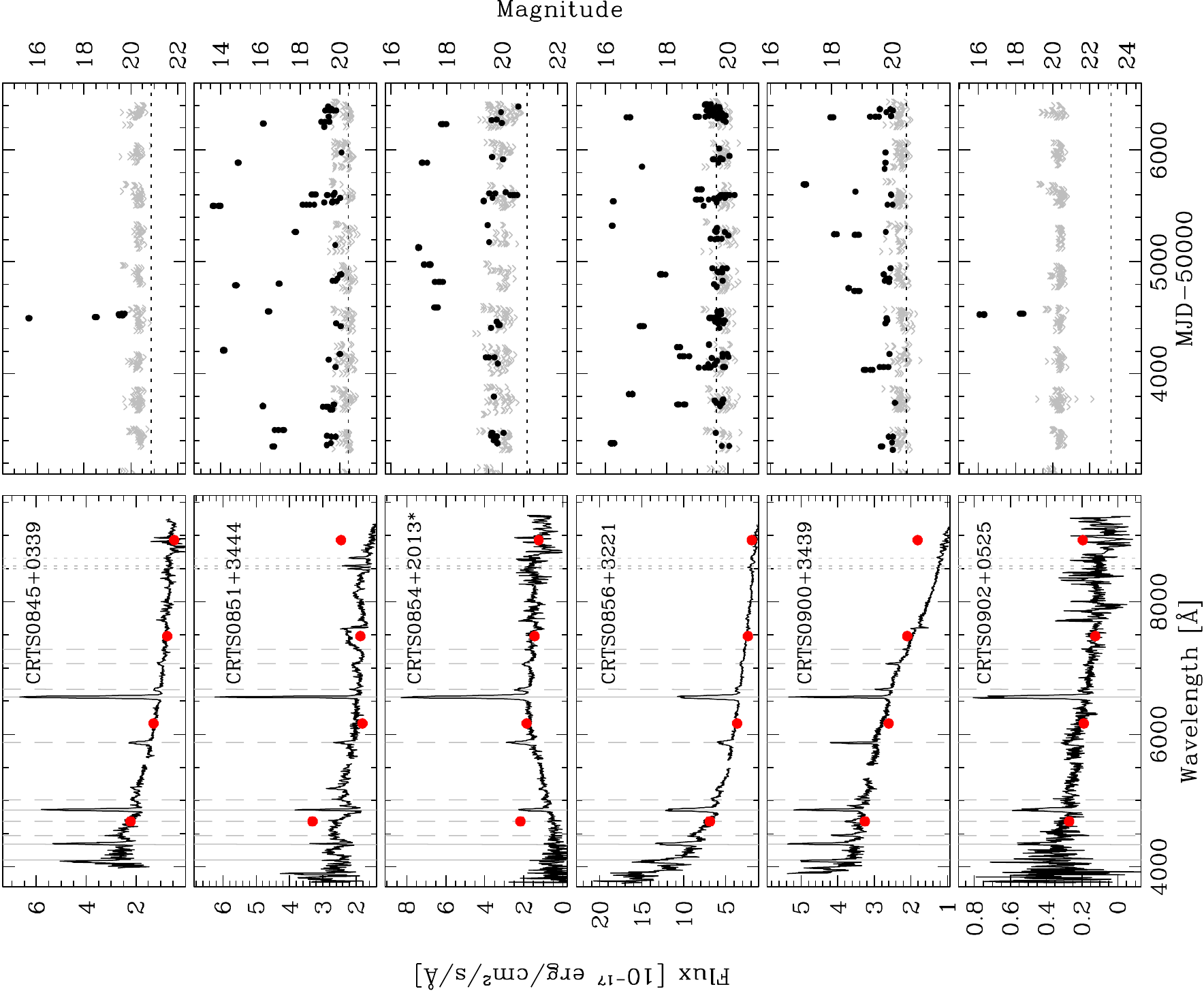}}
\caption{{\em continued.} Spectra and light curves of CRTS CVs.}
\end{figure*}
 
\setcounter{figure}{0}
\begin{figure*}
\centering
\rotatebox{270}{\includegraphics[height=17.5cm]{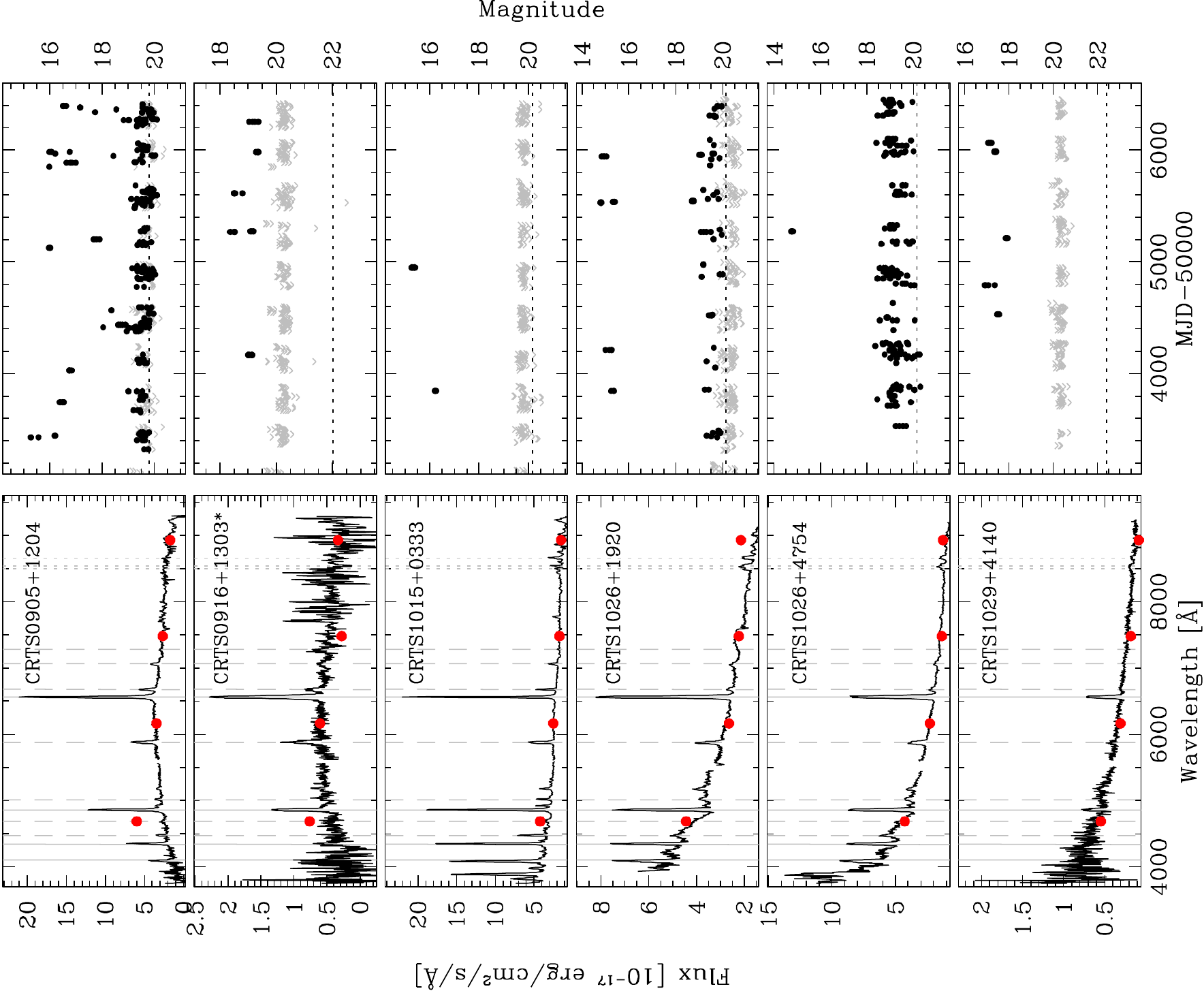}}
\caption{{\em continued.} Spectra and light curves of CRTS CVs.}
\end{figure*}
 
\setcounter{figure}{0}
\begin{figure*}
\centering
\rotatebox{270}{\includegraphics[height=17.5cm]{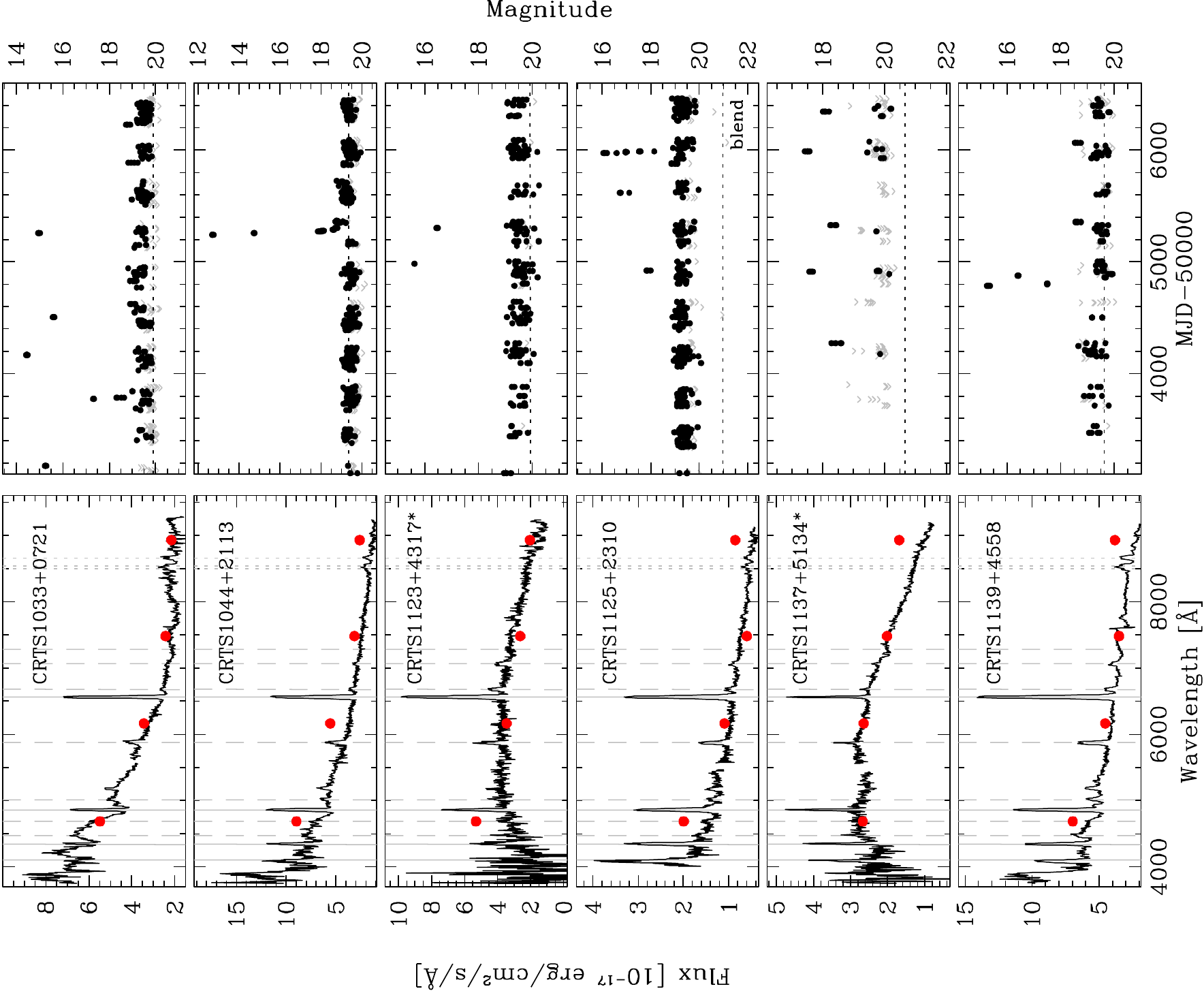}}
\caption{{\em continued.} Spectra and light curves of CRTS CVs. The SDSS image of CRTS\,J112509.7+231036 shows that there is another star very nearby, unresolved in the CRTS images. Judging by the SDSS magnitude (shown as a horizontal dotted line), most of the photometric detections at $g\sim19$ are probably from the unresolved nearby star, and not from CRTS\,J112509.7+231036 itself.}
\end{figure*}
 
\setcounter{figure}{0}
\begin{figure*}
\centering
\rotatebox{270}{\includegraphics[height=17.5cm]{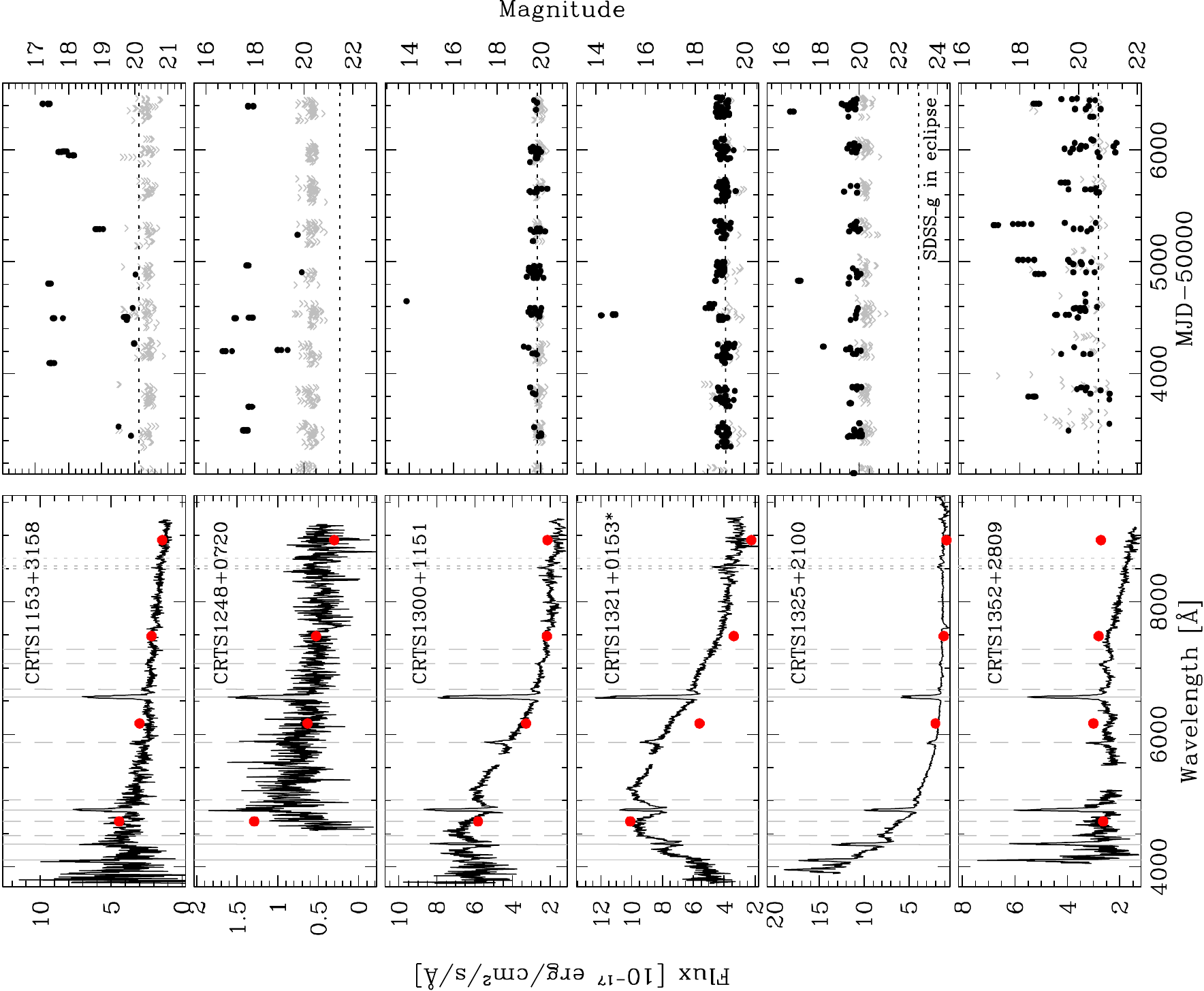}}
\caption{{\em continued.} Spectra and light curves of CRTS CVs. CRTS\,J132536.0+210037 is an eclipsing CV. The $g$ band measurement was taken while the system was in eclipse, so is much fainter than the $riz$ magnitudes and does not appear on the spectrum plot. From the light curve, it is clear that the quiescent brightness is $\sim19.5$ mag.}
\end{figure*}
 
\setcounter{figure}{0}
\begin{figure*}
\centering
\rotatebox{270}{\includegraphics[height=17.5cm]{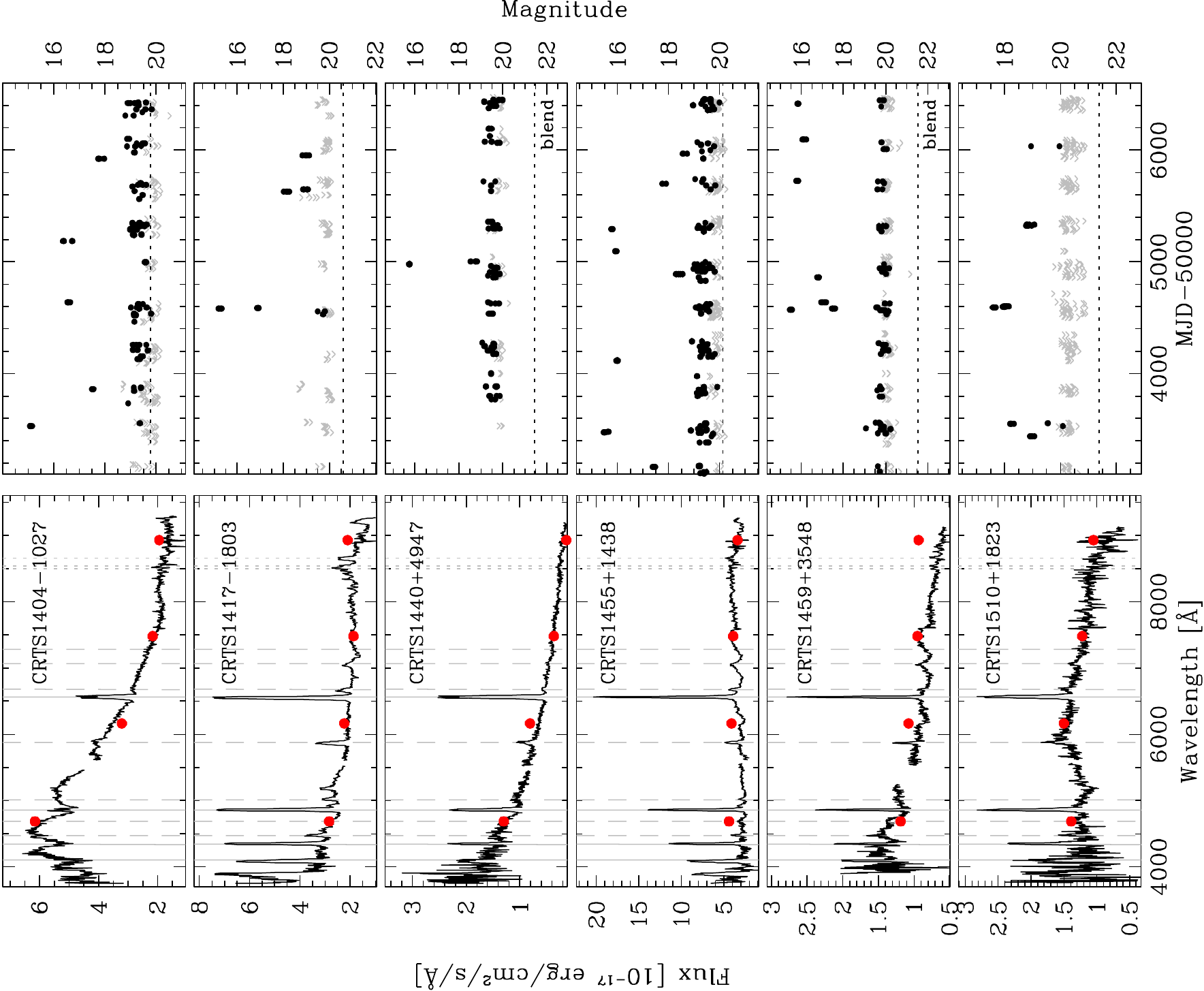}}
\caption{{\em continued.} Spectra and light curves of CRTS CVs. The SDSS images of CRTS\,J144011.0+494734 and CRTS\,J145921.8+354806 show that both targets have another star nearby, unresolved in the CRTS images. It is likely that some of the photometric detections at $g\sim20$ are from the nearby stars rather than from the CVs. The SDSS magnitudes shown refer to the CV only. The stars are resolved in SDSS.}
\end{figure*}
 
\setcounter{figure}{0}
\begin{figure*}
\centering
\rotatebox{270}{\includegraphics[height=17.5cm]{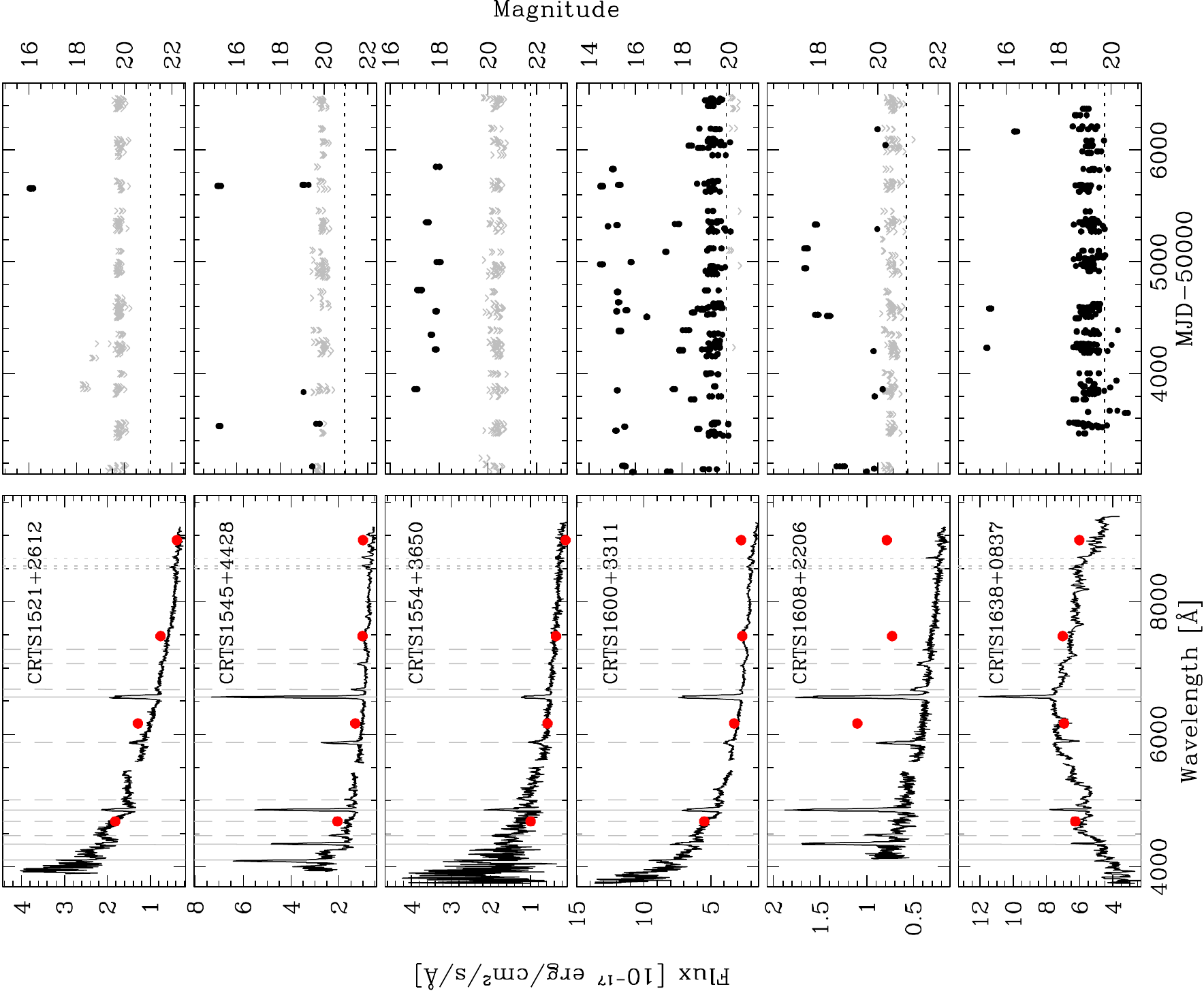}}
\caption{{\em continued.} Spectra and light curves of CRTS CVs.}
\end{figure*}
 
\setcounter{figure}{0}
\begin{figure*}
\centering
\rotatebox{270}{\includegraphics[height=17.5cm]{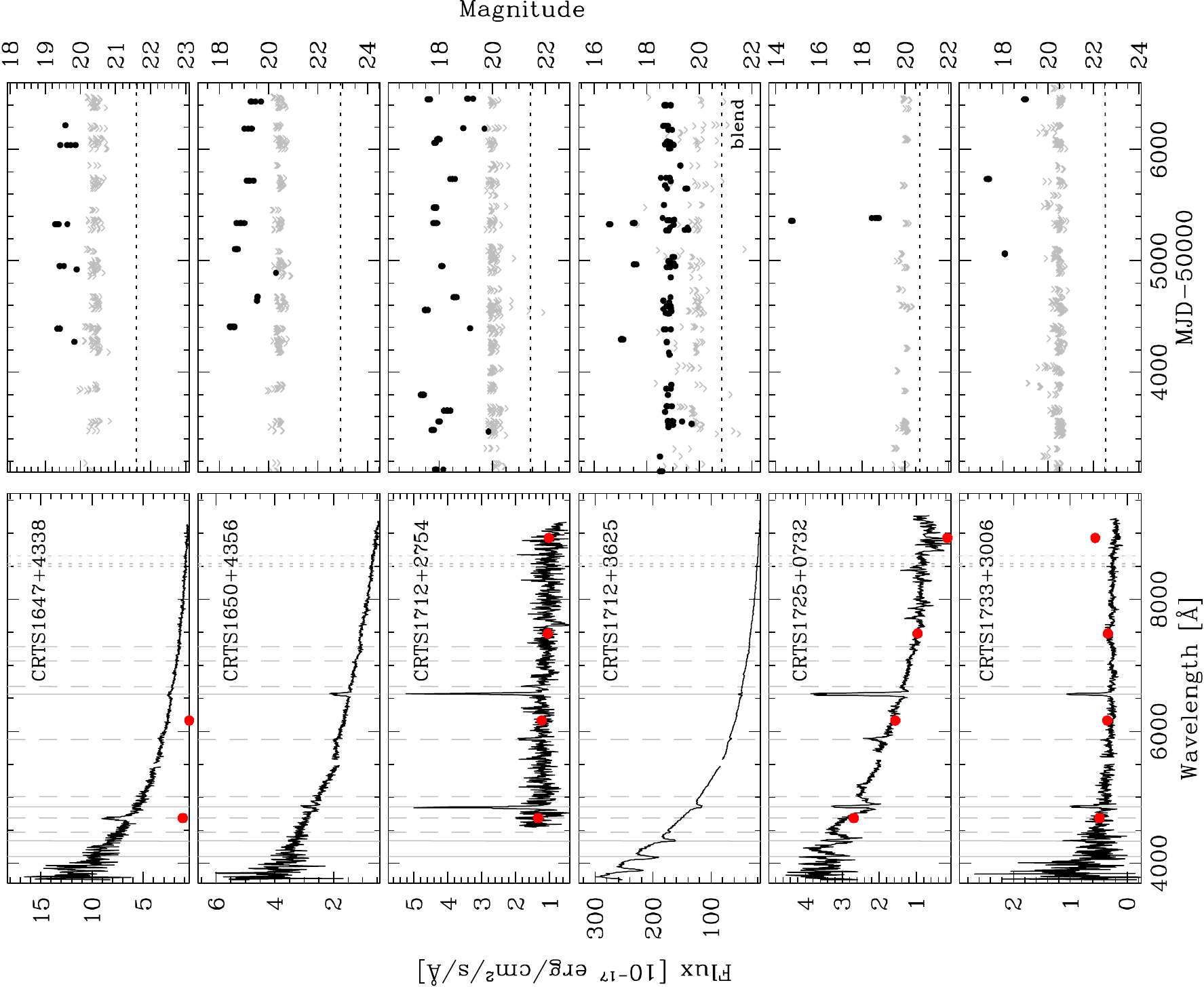}}
\caption{{\em continued.} Spectra and light curves of CRTS CVs. CRTS\,J165002.8+435616 was detected at 2.2 mag brighter than the SDSS photometry at the time of the Gemini observation, similar to the photometric detections shown in the CRTS light curve. CRTS\,J171223.1+362516 was in outburst during our observations, so it was considerably brighter than the SDSS photometry. Additionally, its SDSS image shows that it is blended with a nearby star. Most of the detections at $g\sim19$ are likely to be from the nearby, unresolved star. Its SDSS $g$ band magnitude is $20.87$, as indicated by the dotted horizontal line.}
\end{figure*}
 
\setcounter{figure}{0}
\begin{figure*}
\centering
\rotatebox{270}{\includegraphics[height=17.5cm]{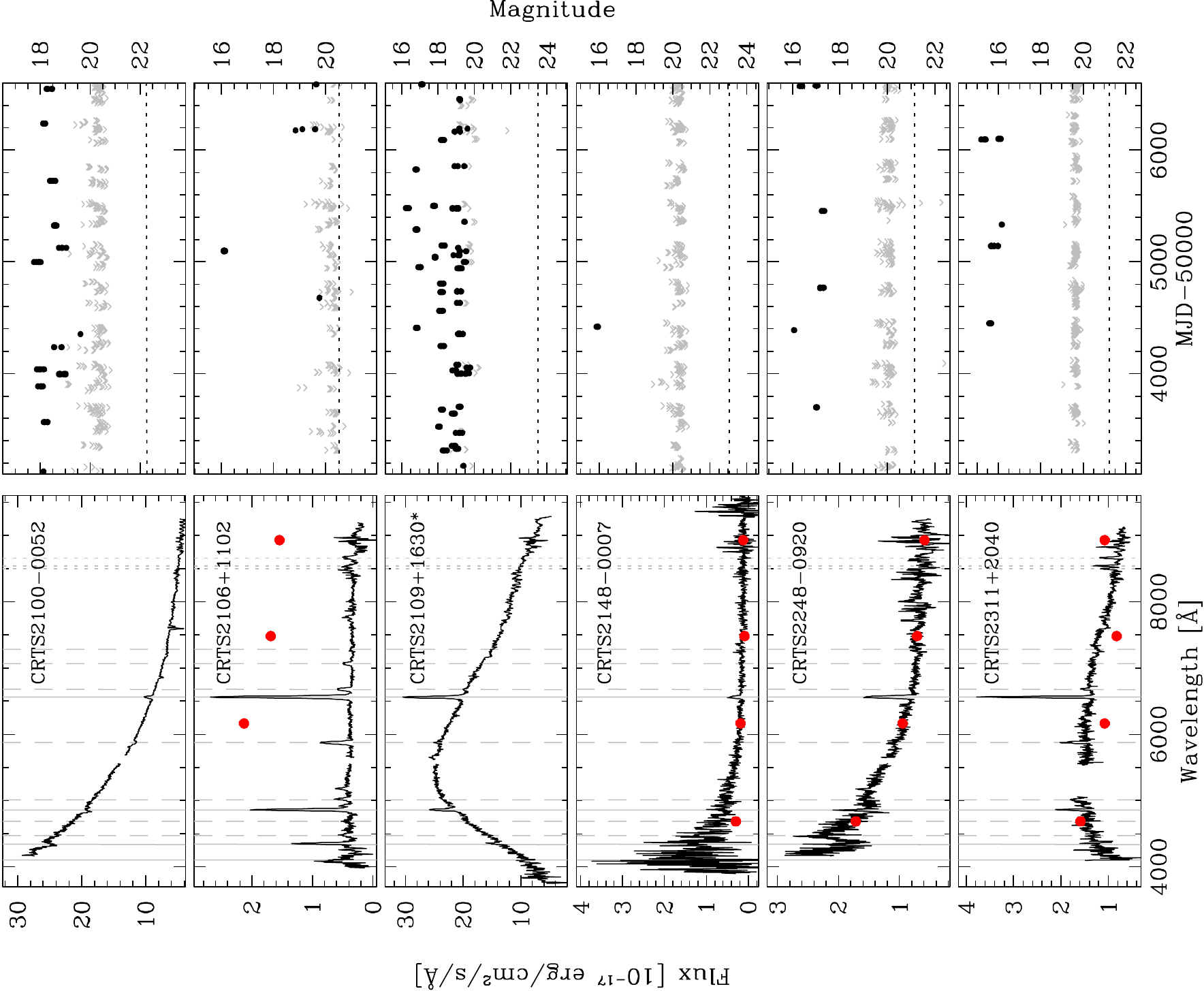}}
\caption{{\em continued.} Spectra and light curves of CRTS CVs. The SDSS photometry of CRTS\,J210650.6+110250 was taken while the system was in a bright state on the rise to or just after an outburst, so do not match the Gemini spectrum. Both CRTS\,J210043.9-005212 and CRTS\,J210954.1+163052 were in outburst at the time of the Gemini observations. CRTS\,J210954.1+163052 is affected by flux loss due to atmospheric dispersion in the blue part of the spectrum.}
\end{figure*}


\label{lastpage}

\end{document}